\documentclass[twocolumn]{aastex631}
\usepackage[utf8]{inputenc}
\usepackage{amsmath}

\shorttitle{Impact of Perturbed Beams on Power Spectrum I}
\shortauthors{Kim et al.}

\renewcommand{\baselinestretch}{1}
\newcommand{\hMpcinv}{h \,\rm Mpc^{-1}}
\newcommand{\kpara}{k_\parallel}
\newcommand{\kperp}{k_\perp}

\begin{document}

\title{The Impact of Beam Variations on Power Spectrum Estimation for 21-cm Cosmology I:\\Simulations of Foreground Contamination for HERA}

\correspondingauthor{Honggeun Kim}
\email{hgkim@mit.edu}

\author[0000-0001-5421-8927]{Honggeun Kim}
\affiliation{Department of Physics, Massachusetts Institute of Technology, Cambridge, MA, USA}
\affiliation{MIT Kavli Institute, Massachusetts Institute of Technology, Cambridge, MA, USA}

\author[0000-0001-5122-9997]{Bang D. Nhan}
\affiliation{Central Development Laboratory, National Radio Astronomy Observatory, Charlottesville, VA, USA}

\author[0000-0002-4117-570X]{Jacqueline N. Hewitt}
\affiliation{Department of Physics, Massachusetts Institute of Technology, Cambridge, MA, USA}
\affiliation{MIT Kavli Institute, Massachusetts Institute of Technology, Cambridge, MA, USA}

\author[0000-0002-8211-1892]{Nicholas S. Kern}
\affiliation{Department of Physics, Massachusetts Institute of Technology, Cambridge, MA, USA}
\affiliation{MIT Kavli Institute, Massachusetts Institute of Technology, Cambridge, MA, USA}

\author[0000-0003-3336-9958]{Joshua S. Dillon}
\affiliation{Department of Astronomy, University of California, Berkeley, CA, USA}

\author[0000-0001-8530-6989]{Eloy de~Lera~Acedo}
\affiliation{Cavendish Astrophysics, University of Cambridge, Cambridge, UK}

\author[0000-0001-7010-0937]{Scott B. C. Dynes}
\affiliation{MIT Kavli Institute, Massachusetts Institute of Technology, Cambridge, MA, USA}

\author[0000-0003-2560-8023]{Nivedita Mahesh}
\affiliation{School of Earth and Space Exploration, Arizona State University, Tempe, AZ, USA}
\affiliation{Cahill Center for Astronomy and Astrophysics, California Institute of Technology, Pasadena, CA, USA}

\author[0000-0001-5300-3166]{Nicolas Fagnoni}
\affiliation{Cavendish Astrophysics, University of Cambridge, Cambridge, UK}

\author[0000-0003-3197-2294]{David R. DeBoer}
\affiliation{Department of Astronomy, University of California, Berkeley, CA, USA}

\begin{abstract}
    Detecting cosmological signals from the Epoch of Reionization (EoR) requires high-precision calibration to isolate the cosmological signals from foreground emission. In radio interferometery, perturbed primary beams of antenna elements can disrupt the precise calibration, which results in contaminating the foreground-free region, or the EoR window, in the cylindrically averaged power spectrum. For Hydrogen Epoch of Reionization Array (HERA), we simulate and characterize the perturbed primary beams induced by feed motions such as axial, lateral, and tilting motions, above the 14-meter dish. To understand the effect of the perturbed beams, visibility measurements are modeled with two different foreground components, point sources and diffuse sources, and we find different feed motions present a different reaction to each type of sky source. HERA's redundant-baseline calibration in the presence of non-redundant antenna beams due to feed motions introduces chromatic errors in gain solutions, which produces foreground power leakage into the EoR window. The observed leakage from vertical feed motions comes predominately from point sources around zenith. Furthermore, the observed leakage from horizontal and tilting feed motion comes predominately from the diffuse components near the horizon. Mitigation of chromatic gain errors will be necessary for robust detection of the EoR signals with minimal foreground bias, and this will be discussed in the subsequent paper.
\end{abstract}

\section{Introduction}
The Epoch of Reionization (EoR), when luminous galaxies formed and interacted with the surrounding intergalactic medium (IGM), is of particular interest for better understanding of the Universe's history. One promising approach to probing this period is to measure 21-cm emission originating from the hyperfine transition of neutral hydrogen at high redshift. This observation is feasible thanks to abundant neutral hydrogen in the early Universe and the transparency of the forbidden transition line \citep[see][for reviews]{Furlanetto2006, McQuinn2016, Liu2020}.

Radio interferometric experiments designed to detect such cosmological signals include the Giant Metre Wave Radio Telescope \citep[GMRT;][]{Paciga2013}, the Murchison Widefield Array \citep[MWA;][]{Tingay2013, Dillon2014, Beardsley2016, Ewall-Wice2016, Barry2019, Trott2020}, the Donald C. Backer Precision Array for Probing the Epoch of Reionization \citep[PAPER;][]{Parsons2010, Cheng2018, Kolopanis2019}, the Low Frequency Array \citep[LOFAR;][]{vanHaarlem2013, Patil2017, Gehlot2019, Mertens2020}, and the Hydrogen Epoch of Reionization Array \citep[HERA;][]{Dillon2016, DeBoer2017, Kern2022}. They have placed upper limits on the power spectrum but none has yet made a robust detection of the EoR signal.

In redshifted 21-cm observations for the EoR, removal of the foreground, which is four to five orders of magnitude brighter, is crucial. One strategy to separate the cosmological signal from the foreground is to take advantage of the two-dimensional power spectrum defined by baseline length and time delay of an interferometer. A radio interferometer measures the sky signal by cross-correlating voltage outputs received by a pair of antennas. Spectral smoothness of foregrounds confines a frequency Fourier transform of the measurements to low time delays or line-of-sight cosmological modes $\kpara$. Chromaticity of the interferometer can introduce spectral structure into the foreground emission, and cause foregrounds to leak into high $\kpara$ \citep{Datta2010, Vedantham2012, Morales2012, Morales2018, Parsons2012, Trott2012, Thyagarajan2013, Liu2014, Pober2014}. The maximal $\kpara$ is set by the delay of a source at the horizon which increases with increasing baseline length. This forms a ``foreground wedge'' in the two-dimensional power spectrum and leaves the foreground confined in the wedge. In contrast, fluctuation in brightness of cosmological signals in the line-of-sight direction or frequency makes the measurement spread over a wide range of cosmological modes in the Fourier space, allowing a foreground-free ``EoR window'' outside the wedge.

For successful foreground removal, high-precision calibration is essential to prevent foreground contamination from leaking into the EoR window due to chromatic errors from inaccurate calibration. There are calibration techniques adopted by past and current interferometric experiments, including sky-based calibration \citep{Pearson1984, Rau2009} and redundant-baseline calibration \citep{Wieringa1992, Liu2010, Dillon2020}. The former relies on precise prior information of the sky, and an inaccurate or incomplete sky model can cause artificial frequency structure in calibration solutions \citep{Barry2016, Ewall-wice2017, Mouri2019, Gehlot2021}. The latter is accompanied by a key assumption of redundancy in measurements, and non-redundancy resulting from non-uniform primary beams can be a source of the chromatic gain error \citep{Orosz2019, Byrne2019, Choudhrui2021}. In addition, redundant calibration cannot solve for all degrees of freedom of the direction-independent gain term, and must be followed by an absolute calibration step, which is also subject to the biases described above for sky calibration \citep{Byrne2019, Kern2020a}.
In this study, we focus on the impact of feed motion on redundant-baseline calibration as well as its absolute calibration step as designed for the HERA instrument.

\citet{Orosz2019} studied the impact of non-uniform primary beams on redundant-baseline calibration and resulting power spectrum estimate. They simulated interferometric measurements with the HERA array configuration and an analytic Airy beam by perturbing the beam pointing angle and width. They found perturbed primary beams can introduce spectral structure into calibration solutions which are then responsible for the foreground in the wedge leaking into the EoR window. The actual HERA antennas have more complex structure in their beams and the result based on the Airy beam is hard to map to the real system of HERA. In addition, the simulation was performed with about 100 bright point sources and the effect of a diffuse sky model on the power spectrum analysis was not addressed. These motivate us to take one step further with a physical antenna model and a more representative sky model. \citet{Shaw2014} also studied the impact of per-antenna primary beam deviations on foreground removal for low-frequency 21-cm intensity mapping, finding that per-antenna beam deviations need to be constrained below 10\% for robust signal detection. 

In this study, we employ a realistic primary beam obtained from a physical HERA antenna model using the CST electromagnetic simulation software. In the model, we perturb positions of the antenna feed relative to the dish and compute the corresponding primary beam patterns. We focus on the HERA Phase II system with a new broadband (50--250~MHz) Vivaldi feed that is designed to cover the epoch of Comic Dawn at lower frequencies as well as the EoR \citep{Fagnoni2021}. We choose a mid frequency band, 160--180~MHz, corresponding to $z \sim 7.4$ that is a midpoint of the late EoR model \citep{Mitra2015, Doupis2015, Greig2017, Millea2018, Qin2020} consistent with the constraints on the electron scattering optical depth of the Cosmic Microwave Background observation \citep{Plank2018}. For a representative foreground model, a diffuse sky model as well as point sources are taken into account. We simulate visibilities with the perturbed primary beams and the sky model, and pass them through HERA calibration pipelines. The effects of non-redundancy in visibilities on the calibration gains and power spectrum estimation are explored. Further investigation into mitigation of the chromatic gain errors and foreground leakage will be discussed in the subsequent paper.

This paper is organized as follows. In Section~\ref{sec:beam_sim}, we detail the configuration of primary beam simulations with feed motions and present characteristics of the simulated beams. In Section~\ref{sec:vis_sim}, visibility measurements are simulated with the perturbed beams, and non-redundancy of the visibilities are investigated as a function of baseline lengths, feed motions, and the type of sky model. In Section~\ref{sec:calibration}, we show results from redundant-baseline calibration as well as absolute calibration, representing chromatic gain errors. In Section~\ref{sec:pspec}, the effects of the feed motions on power spectrum estimation are discussed. Throughout this study, we adopt the cosmological parameters from \citet{Planck2016}: $\Omega_\Lambda = 0.6844$, $\Omega_{\rm m} = 0.3156$, $\Omega_{\rm b} = 0.04911$, and $H_0 = 67.27\,{\rm km\, s^{-1}\, Mpc^{-1}}$.

\section{Antenna Feed Perturbation Simulations}
\label{sec:beam_sim}
In this section we summarize the CST simulation descriptions and results of the antenna far-field electric field pattern in the presence of positional offsets between the Vivaldi feed and the 14-meter parabolic dish. The physical three-dimensional model was derived from that of \cite{Fagnoni2021}. The model consists of all the major components making up a single HERA antenna, including metal, PVC, and fiberglass components, coaxial cables, an enclosure for the front-end module (FEM) with its CAT7 connector, a cylindrical ferrite radio frequency (RF) choke for each coaxial cable, metal mounting hardware, suspension cables, a concrete hub, and a soil slab\footnote{The material is assumed to be sandy dry soil with relative permitivity of $\epsilon_r=2.55$ and loss tangent of $\tan\delta = 0.016$.} beneath the dish.

\subsection{CST Simulation Configuration}
\label{sec:cst_settings}
Due to wind load and relaxation in the feed suspension system, the feed alignment can move away from the desired feed position at the focal point of the dish. Field measurement of the feed alignment using laser plum bobs on a subset of HERA elements across a time span of 8 weeks in 2020 determined that the standard deviation of the feed offset can be as large as several centimeters (Rath et al. HERA Memo \#95\footnote{\url{http://reionization.org/science/memos/}\label{footnote1}}). Separate observations on tilts of feeds measured by accelerometers installed in the FEM indicate a root mean square (RMS) of the tip-tilt of a few degrees.

Due to computational limitations, a simplified three-dimensional Vivaldi feed model was derived from the detailed model developed by \citet{Fagnoni2021}. We removed the smaller mechanical features, such as screw mounting holes, coaxial cables, suspension cables, and the FEM. For a subset of the feed motions, we found that the overall far-field beam characteristics for the simplified model and the detailed model are consistent, and the simplification does not have a large impact on our study. The primary beam of each antenna element is computed in an isolated simulation domain bounding box and we ignore mutual interactions between antennas such as a cross-coupling effect in the HERA system. However, we expect that cross-coupling might degrade calibration solutions \citep[see, for example,][]{Kern2020b, Josaitis2022}, and calibration procedures that include a model of cross-coupling are under development. We defer a quantitative study of the interaction of beam imperfections and cross-coupling 
to future work.

\begin{figure*}[t!]
\centering
\includegraphics[scale=0.42]{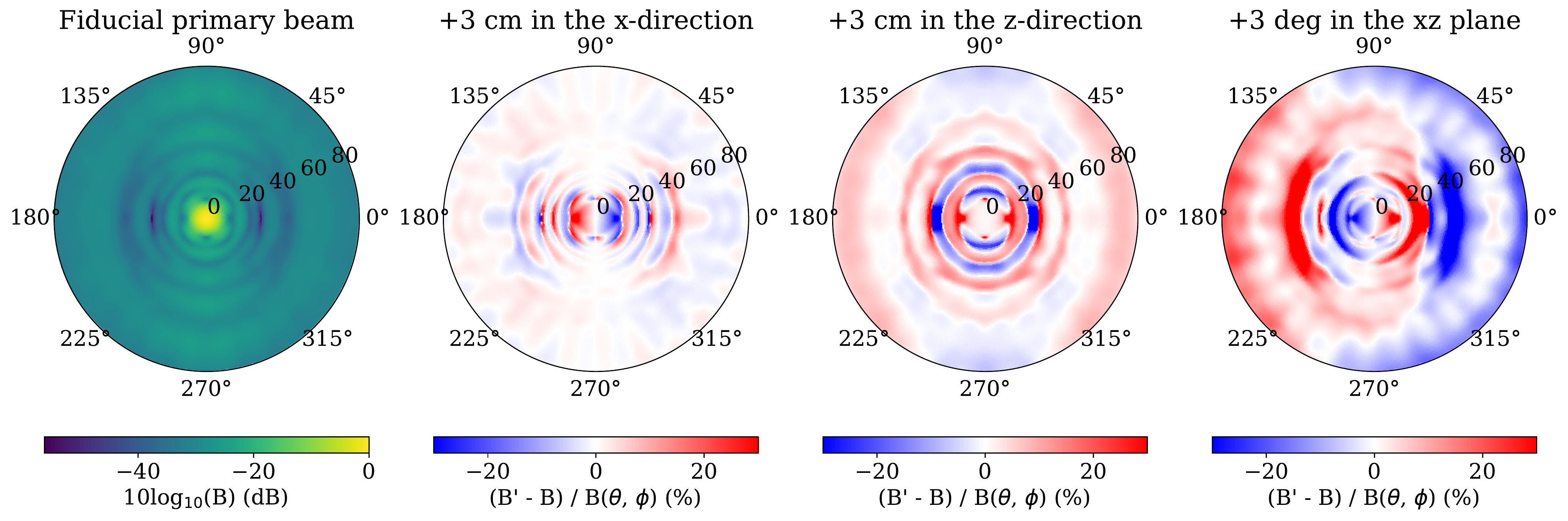}
\caption{The fiducial beam pattern (far left) and the difference of the primary beam models (B') relative to the fiducial model (B) for feed displacement by +3 cm in the $x$-direction (center left), by +3 cm in the $z$-direction (center right), and tilt by $+3^{\circ}$ in the $xz$-plane (far right) at 165 MHz. The difference is normalized by the amplitude of the fiducial model. The different feed motion leads to the different beam pattern.}
\label{fig:beam_diff}
\end{figure*}

The antenna dish model consists of 24 faceted parabolic panels as in the deployed instrument. However, instead of using actual galvanized steel wire mesh panels, the dish model assumes  solid aluminum panels. This simplification circumvents the challenge in properly simulating the wire mesh and significantly reduce the number of simulation meshcells. Fabrication imperfections such as surface smoothness, potential gaps between dish panels, or structural sagging due to gravity are not considered in this study. Due to the randomness in nature, those defects may lower the antenna gain due to signal loss but they are expected to have much less impact on the overall main beam's characteristics than the feed position perturbations.

The Vivaldi feed position was perturbed in the lateral $xy$-plane and along the boresight direction in the $z$-axis, separately, in the CST model. We set the range of feed offsets along the $z$-axis to be $\pm$7~cm, with increment of 0.5~cm between -2 and 2~cm and of 1~cm otherwise, relative to the fiducial feed height position of $(x,y,z) = (0.0, 0.0, 5.0)$~m where the $z$-position is the distance between the reference point on the feed \citep[see the origin in Figure~2 of][]{Fagnoni2021} and the vertex of the dish. The offsets in the $xy$-plane spanning $\pm$6~cm are sampled in a regular grid with 0.5~cm interval between -2 and 2~cm and 1~cm interval otherwise. We include an additional 9 points at the corner and the midpoint of each side of a 7-cm long square to deal with potential outliers.

The feed can also tilt relative to the dish as the suspending cables relax over time. We simulated tilts at the fiducial position only, without any additional transverse offset. The feed is tilted in the range of $\pm 6^{\circ}$, with increment of $0.5^{\circ}$ between $\pm 1^{\circ}$ and of $1^{\circ}$ otherwise, where $0^{\circ}$ tilt is equivalent to the feed pointing straight down at the dish's vertex. The azimuthal motion is equally spaced at 10$^\circ$ for each tilt at the fiducial position.

In total, we have three different classes of the feed motions for the CST beam simulations: one set of CST simulations produced 19 perturbed beam patterns for the vertical displacement, another set produced 209 perturbed beams for all the horizontal offsets in the $xy$ plane, and the other set produced 295 perturbed beam patterns for all the tilts at the fiducial position.

Far-field electric fields are simulated by exciting the East-West port over 160--180~MHz, with a frequency channel width of 0.125~MHz\footnote{The far-field results were exported with high precision to prevent round-off errors using CST's far-field source (FFS) export instead of the normal far-field export macro function. This is essential to achieve a large dynamic range in power spectrum estimation.}. This frequency channel width is close to that of the HERA Phase II observation, and gives the large Nyquist limit needed for quantifying the amount of contamination in the EoR window at high $\kpara$.

\begin{figure*}[t!]
\centering
\includegraphics[scale=0.42]{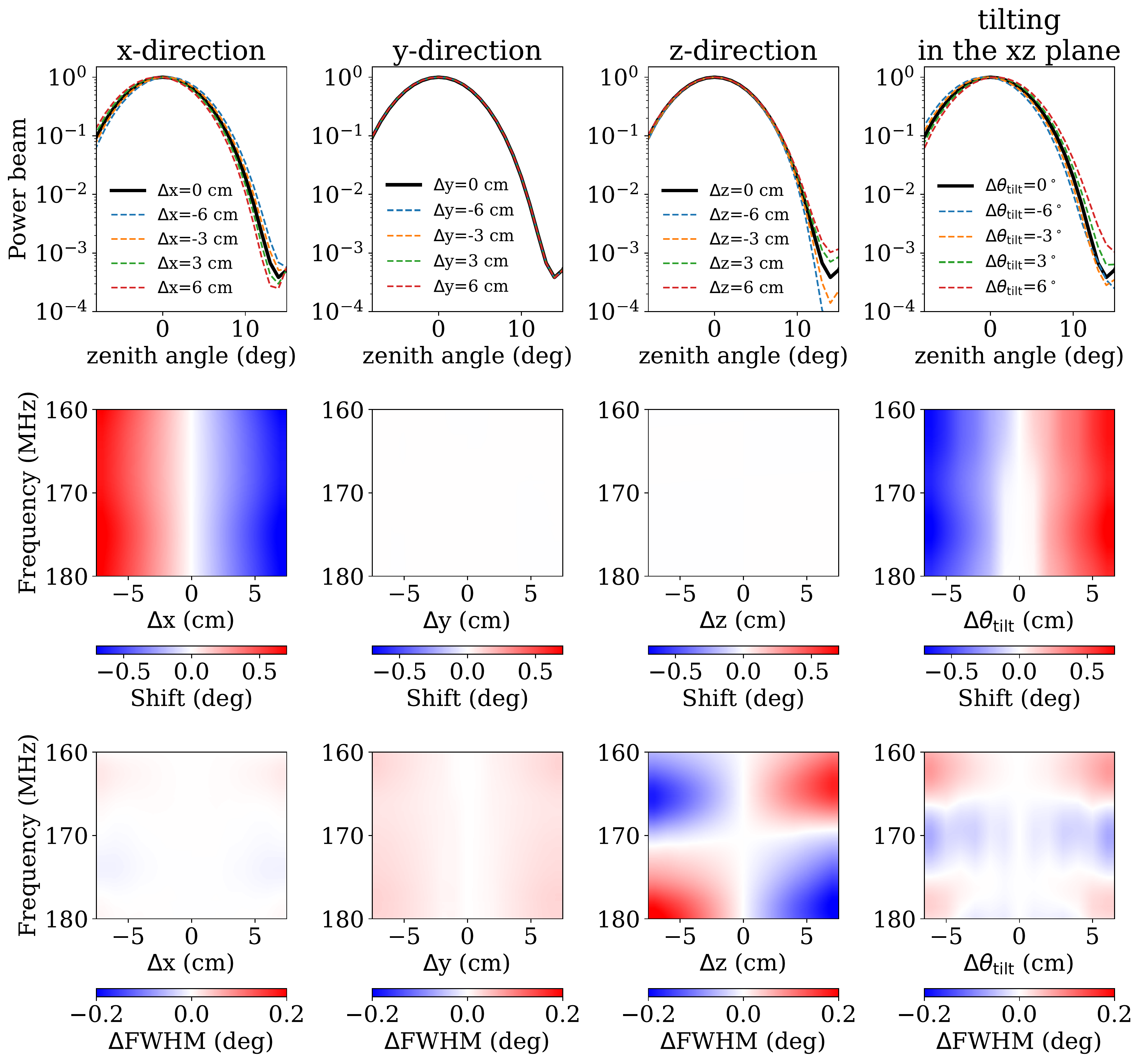}
\caption{Behaviors of the main lobe of the power beam in the $xz$ plane or in the EW direction with different types of feed motions: $x$-direction (far left), $y$-direction (center left), $z$-direction (center right), and tilting feed motions in the $xz$ plane (far right). The top row shows line profiles of power beams that are normalized to the peaks at 165 MHz. Based on line profiles, we measured the shift of maximum of the main lobe (middle row) and the FWHM of the main lobe (bottom row) as a function of feed motion and frequency. When the feed moves in $x$-direction or tilts in the $xz$ plane, the main lobe shifts in the $xz$ plane as expected. The beam width changes relatively more when the feed moves in the $z$-direction.}
\label{fig:main_lobe_xz_plane}
\end{figure*}

The full-wave far-field simulation results were computed using CST's transient time solver based on Finite-Difference Time-Domain (FDTD) solver method with open boundary conditions\footnote{The perfectly matched layer (PML) boundary conditions were applied at the the boundary interfaces of the simulation bounding box to emulate an infinite simulation domain for the electromagnetic wave to propagate outward with negligible reflection at the interfaces.}. Benchmark tests suggest that CST's time solver with GPU acceleration\footnote{Combining with local meshing scheme for small antenna components, the simplified antenna model with $\sim 7.5\times10^7$ volumetric meshcells took an average of 3--5 hours to simulate and export each beam set on a computer server equipped with two Tesla V100-PCIE-32GB GPUs.} is more efficient for the required large number of beam simulations. A subset of the far-field results was compared with those of CST's Finite Element Method frequency solver, which is not compatible with GPU acceleration. The general beam characteristics due to feed motions are found to be consistent between the two solvers.

\begin{figure*}[t!]
\centering
\includegraphics[scale=0.42]{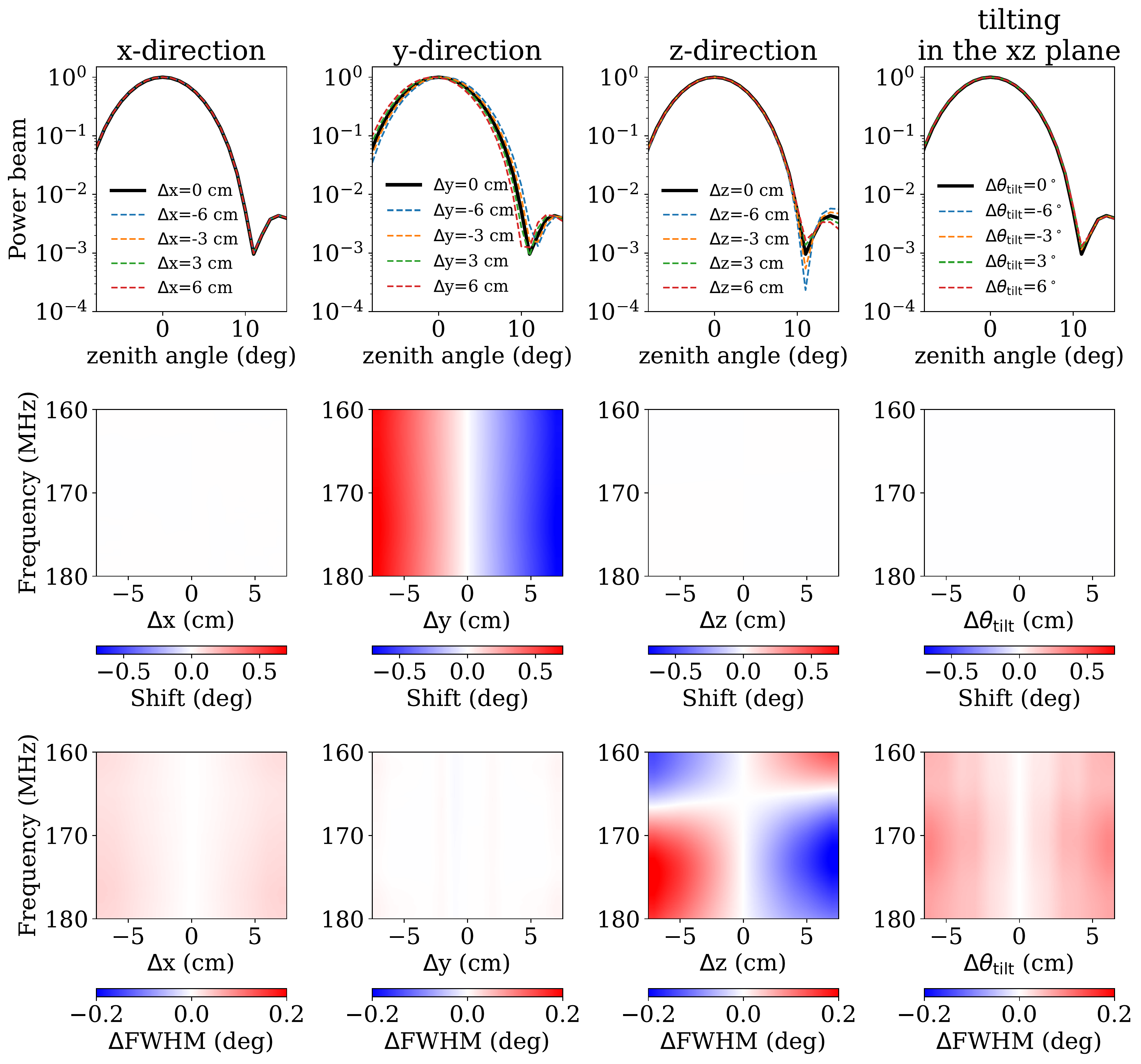}
\caption{The same format as that of Figure~\ref{fig:main_lobe_xz_plane} but for the behaviors of the primary beam in the $yz$ plane or in the NS direction. The shift of the peak location in the main lobe is observed for the $y$-direction feed motion, while the change to the beam width mainly arises from the vertical motion or the tilting motion.}
\label{fig:main_lobe_yz_plane}
\end{figure*}

\subsection{Characteristics of Primary Beams along the Feed Motion}
\label{sec:characteristics_beams}
Feeds which move away from their fiducial position produce perturbed far-field electric fields which are different from those at the fiducial position. This perturbed primary beam pattern can corrupt interferometric measurements and calibration pipelines. In this section, we present characteristics of the primary beams as a function of feed positions and frequency. We examine the shift and full width at half maximum (FWHM) of the main lobe of the power beam with feed motion along the three Cartesian axes or with tilt. In addition, we explore the spatial integral of the perturbed beam with respect to the unperturbed one to capture overall features of the perturbed beams with feed motion and frequency.

The far-field electric fields simulated by CST of a linear feed polarization $p$ are defined in a basis of unit vectors $\hat{\theta}$ and $\hat{\phi}$ in spherical coordinate,
\begin{align}
    \mathbf{E}^p(\theta, \phi, \nu) = E^p_\theta(\theta, \phi, \nu)\hat{\theta} + E^p_\phi(\theta, \phi, \nu)\hat{\phi},
    \label{eqn:efield}
\end{align}
where $\theta$ and $\phi$ are the zenith angle and the azimuthal angle, respectively. \autoref{eqn:efield} can also be thought of as a row in the antenna Jones matrix \citep[e.g., Equation~5 of][]{Kohn2019}. The power beams are then calculated by multiplying each component of far-field electric fields assuming unpolarized sky emissions \citep{Kohn2019},
\begin{align}
    B^{pp}_{ij}(\theta, \phi, \nu) = E^p_{i,\theta} {E^p_{j,\theta}}^* + E^p_{i,\phi} {E^p_{j,\phi}}^*,
    \label{eqn:power_beam}
\end{align}
where $pp$ is a feed polarization pair, and $i$ and $j$ indicate a pair of antennas. Throughout this paper, we consider a single feed polarization (``East-oriented'') in our CST and visibility simulations.

Figure~\ref{fig:beam_diff} displays the fiducial power beam (far left) and the difference between the perturbed feed and the fiducial one at 165 MHz. The feed moving horizontally (center left) or tilting (far right) yields a shift of the beam pattern with that from the tilting motion moving opposite to that from the $x$-direction motion. The feed moving vertically changes the beam width, producing relatively isotropic differences (center right).

\begin{figure*}[t!]
\centering
\includegraphics[scale=0.5]{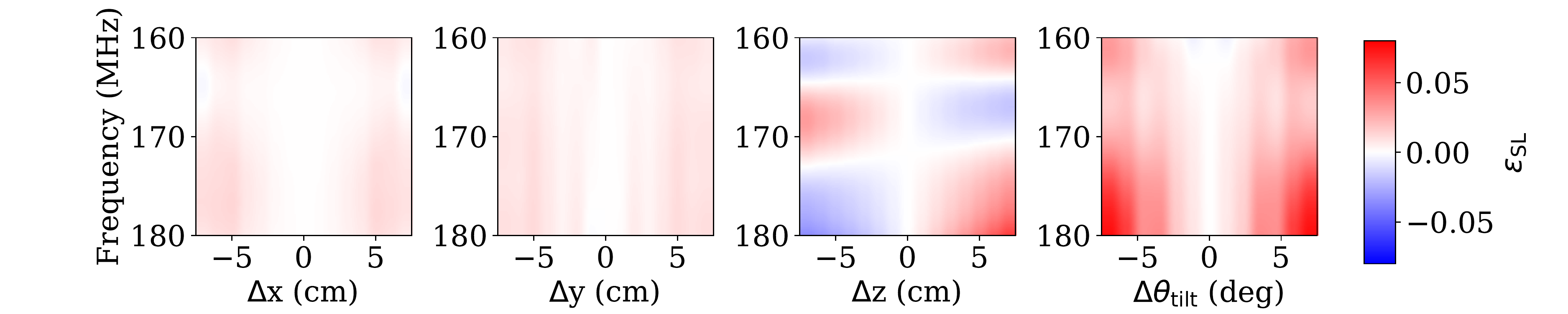}
\caption{Integrated primary beams outside the main lobes of the perturbed beam with respect to the unperturbed beam. For the translational motions shown in the first three panels, the feed displacement is labeled from -7 to 7 cm. For the tilting motion, we consider the motion in the $xz$ plane with tilt measured from zenith spanning from -6$^\circ$ to 6$^\circ$.}
\label{fig:side_lobe_int}
\end{figure*}

To quantify the characteristics of the main lobe, we first locate the maximum of the power beam and fit a one-dimension Gaussian profile to the first null in the $xz$ plane (or in East-West direction, EW) or in the $yz$ plane (or in North-South direction, NS) at boresight of the beam to measure the peak location and the beam width of the main lobe.

Figure~\ref{fig:main_lobe_xz_plane} shows the behaviors of the main lobe of the primary beam with the feed motion. In the top row, we show line-cut profiles of main lobes in the $xz$ plane for the $x$-direction (first column), $y$-direction (second column), $z$-direction (third column), and tilting (last column) feed motion. If we look at the first panel, the line profile of the power beam shifts in the $xz$ plane at different feed positions moving along the $x$-axis. The measured shift of the peak of the main lobe in the $xz$ plane is presented in the middle row as a function of feed offset and frequency. In the first panel of the middle row, there are three interesting things to note. First, the direction of the beam shift is opposite to that of the feed displacement. Second, the shifts of the peaks are frequency independent. Finally, there is a nearly linear relationship between the feed offsets and the beam shifts. These characteristics can be understood from geometrical optics, $\theta \approx -\Delta x/h \,\,  {\rm [rad]} = -0.11 \, (\Delta x / {\rm cm}) \,\, {\rm [deg]}$, where $\Delta x$ is the offset of an object from the fiducial point, and $h$ is the normal distance between the object and a lens. In the second equality, we consider our feed model where $\Delta x$ is the offset of the feed in the $x$-direction and $h = 500$~cm, resulting in the relation consistent with the trend we observe. When the feed moves along the $x$-axis, the FWHM of the main lobe is weakly dependent on the feed motion (far left panel of the bottom row of Figure~\ref{fig:main_lobe_xz_plane}) and this may be because of a defocusing effect when the phase center deviates from the focal point with the feed motion. If the feed moves in the $y$-direction (center left panels of Figure~\ref{fig:main_lobe_xz_plane}), the profile remains nearly the same in the $xz$ plane as expected. 

The change in the beam shape with the feed offset in the $z$-axis is different from that with the lateral feed motion. Unlike the lateral motion, the peak of the main lobe stays unchanged with the feed motion (the third column of Figure~\ref{fig:main_lobe_xz_plane}), because change in the beam shape is symmetric and thus the beam is still pointing along the $z$-axis.. Rather, the axial motion leads to the change of the beam width as shown in the center right of the bottom row. This may be associated with the relation between the phase center and the focal point. That is, when the phase center, moving with the feed height, gets farther away from the focal point, the beam goes through the defocusing effect and its width grows broader \citep{Baars2007}. The phase center is frequency-dependent, so this effect is different for different frequencies.

When the feed tilts, the beam is also expected to tilt in the same direction. In the last column of Figure~\ref{fig:main_lobe_xz_plane}, the primary beam shifts in the EW direction when the feed tilts in the $xz$ plane. Because the main lobe tilts with the feed motion, the beam does not just shift but its width also changes as a function of tilting angle.

In Figure~\ref{fig:main_lobe_yz_plane}, we show the same feed motions but the line-cut profiles in the $yz$ plane. Feed offsets in the $y$-direction cause the shifts of the main lobe in the $yz$ plane, while the feed motions along the $x$-direction or the tilting motions in the $xz$ plane result in consistent line profiles regardless of the size of the feed displacement. The peak location of the main lobe still stays unchanged with the vertical feed motion while there is a clear trend of the change to the beam width as a function of the feed motion and frequency. The FWHM of the main lobe becomes slightly broader in the NS direction with the tilting motion.

In general, the amplitudes of the side lobes are a few order of magnitudes smaller than that of the main lobe. However, integrated properties of the side lobes can be crucial when a sky model contains diffuse bright sources in the side lobes. To measure the variation of side lobes due to feed motion compared to the fiducial model, we define
\begin{align}
    \varepsilon_{\rm SL} &= \frac{\iint B_{\rm pert}(\theta,\phi)\,d\Omega}{\iint B_{\rm unpert}(\theta,\phi)\,d\Omega} - 1.
    \label{eqn:side_lobe_int}
\end{align}
Here $B_{\rm unpert}(\theta,\phi)$ and $B_{\rm pert}(\theta,\phi)$ are unperturbed and perturbed power beam, respectively. The integral is over $60^{\circ} \le \theta <90^{\circ}$ and $0^{\circ} \le \phi < 360^{\circ}$. Anticipating our visibility simulations, the integral range is chosen so that it includes the region where the galactic plane lies at a local sidereal time (LST) of interest, 2.25 hours. The result is shown in Figure~\ref{fig:side_lobe_int}. The largest change in $\varepsilon_{\rm SL}$ is observed for the tilting motion and there is little change for the lateral feed motions.

\section{Visibility Simulations with Perturbed Primary Beams}
\label{sec:vis_sim}
\subsection{Foreground Visibility Simulations}
\label{sec:foreground_sim}
A visibility for two antennas is the interferometric response with the amplitude proportional to the beam power pattern along with the flux density of the sky and the phase depending on the frequency and the geometric delay. This can be described in a discretized form as
\begin{equation}
V_{ij}(\nu) = \sum^{N_{\rm src}}_{n=1} B_{ij}(\hat{s}_n, \nu) S(\hat{s}_n, \nu) \exp\bigg({-\frac{2\pi i\nu}{c} \mathbf{b}_{ij} \cdot \hat{s}_n}\bigg),
\label{eqn:vis_sim}
\end{equation}
\noindent where $B_{ij}$ is the primary beam in Equation~\eqref{eqn:power_beam}, $S$ is the flux density of a source, $\textbf{b}_{ij}$ is a baseline vector of two antennas, and $\hat{s}_n$ is a pointing vector in the direction of a source.

\begin{figure}[t!]
\centering
\includegraphics[scale=0.35]{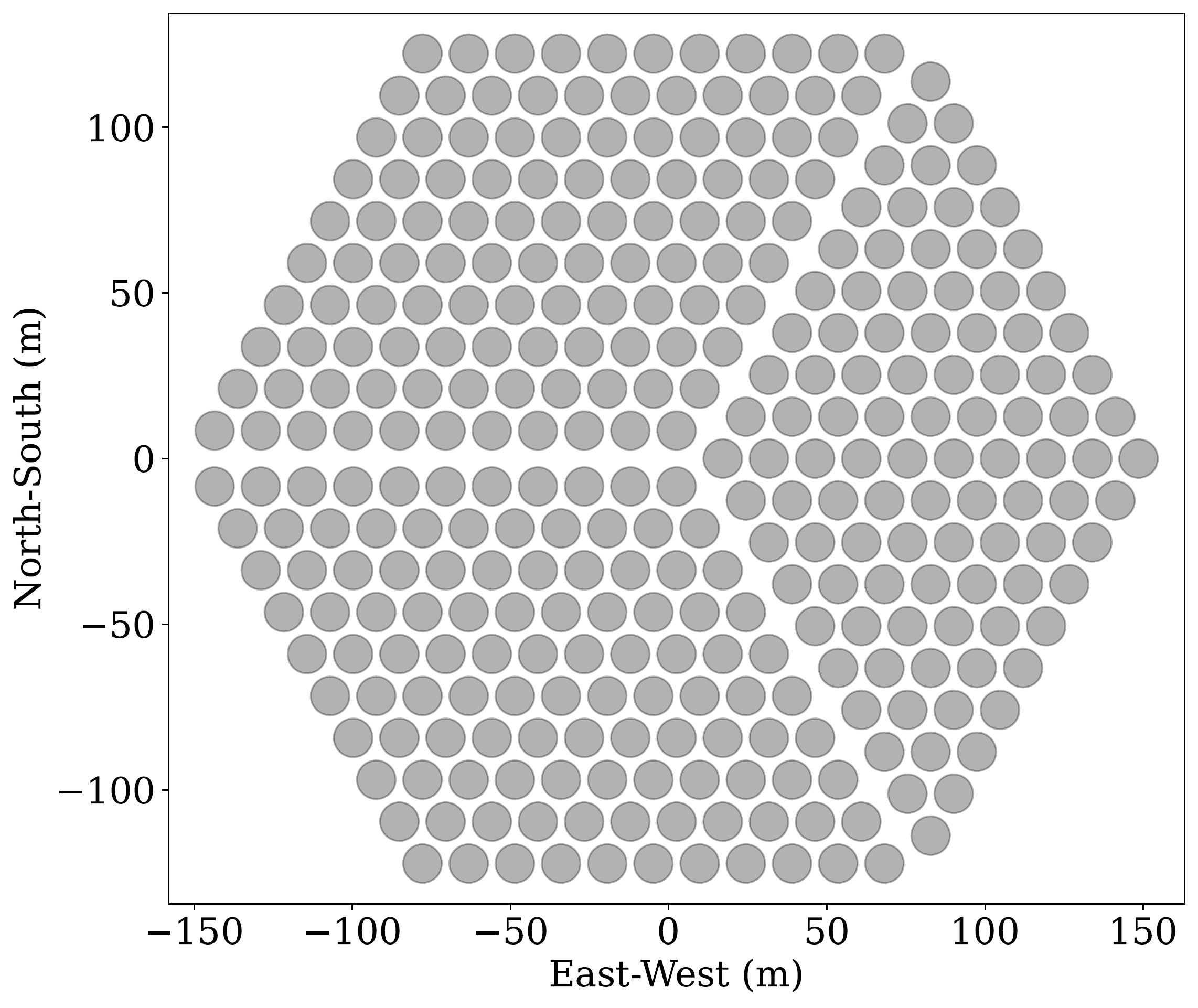}
\caption{Configuration of HERA-320. This highly redundant array configuration is designed to achieve high-precision calibration and deep sensitivity to the 21-cm power spectrum.}
\label{fig:H320}
\end{figure}

In this study, we simulate visiblities using 320 antennas which make up the core of the HERA array \citep[HERA-320,][]{Dillon2016}. The configuration of the array is shown in Figure~\ref{fig:H320}. There are 51,360 baselines in total and 1,502 unique baselines. This highly redundant-baseline configuration enables us to achieve high precision calibration with HERA if all antennas have uniform antenna responses. However, the distinct antenna response among the array elements due to feed displacement invalidates the redundancy assumption thus compromising the calibration. We account for this by assigning a different primary beam to each antenna in Equation~\eqref{eqn:vis_sim} and seeing the effect of non-uniform antenna responses on the calibration.

In practice, the feed position of each antenna can be perturbed in any direction with random distances with respect to the fiducial position. As mentioned in Section~\ref{sec:cst_settings}, the sizes of feed offsets are expected to be a few centimeters for translation motions and a few degrees for tilts. For simplicity, we consider random Gaussian feed motions for 320 antennas with zero mean and a standard deviation ($\sigma_{\rm feed}$) of 1, 2, 3, and 4 cm for lateral and axial feed motions, respectively, and of 1, 2$^\circ$, and 3$^\circ$ for tilting motions.

Since the primary beams are simulated at grid points of feed motions described in Section~\ref{sec:cst_settings}, the primary beam patterns used in Equation~\eqref{eqn:vis_sim} need to be interpolated at the feed positions drawn from the Gaussian distribution. Interpolating the primary beams along the feed motion direction can introduce spectral artifacts. We found, however, such an error is a much smaller effect than the beam variations due to the feed motion. Appendix~\ref{sec:beam_interpolation} describes the interpolation and presents the interpolation error relative to the beam error driven by the feed motion.

In addition to the interpolation along the feed motion direction, the primary beam needs to be evaluated at the frequency of interest and angular positions of sources, requiring additional interpolation. Interpolating the primary beam along frequency at a finer resolution than the beam resolution may introduce unwanted spectral structure which depends on the interpolation methods \citep{Lanman2020}. To avoid this effect, the visibility needs to be simulated with a frequency channel width larger than or equal to the frequency resolution of the primary beam simulation. Large frequency channel widths, however, will shrink the size of the observable EoR window which is proportional to $1/(2\Delta\nu)$ where $\Delta\nu$ is the channel width. Based on the result of \citet{Orosz2019}, in order to quantify the foreground contamination in the EoR window, the frequency resolution of our primary beam, 0.125~MHz, is fine enough to cover a large range of $\kpara$ and thus used for the simulation instead of interpolation along frequency.

\begin{figure*}[t!]
\centering
\includegraphics[scale=0.6]{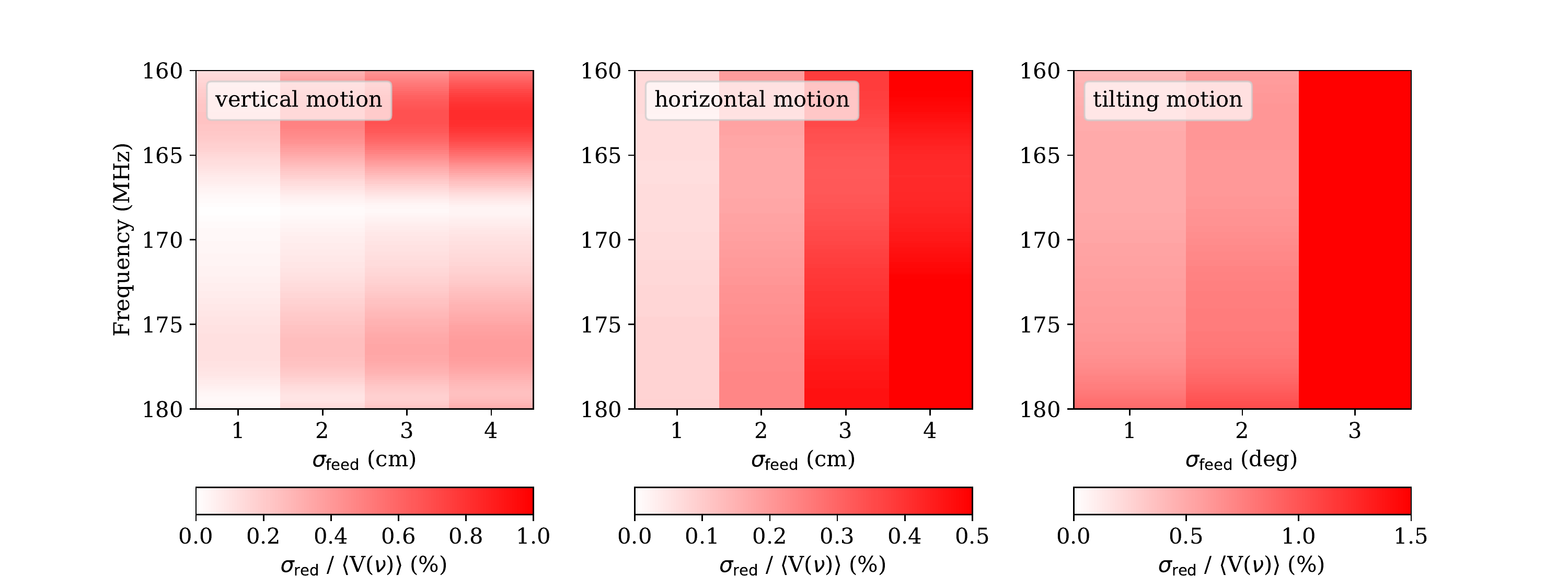}
\caption{Relative $\sigma_{\rm red}$ of auto-correlations for vertical (left), horizontal (middle), and tilting feed motions (right) with feed offset and frequency.}
\label{fig:std_auto}
\end{figure*}

The CST simulated beams are resolved at 1$^\circ$ in the spherical coordinate along $\theta$ and $\phi$. Based on the sampled grid points, we interpolated far-field electric fields defined in Equation~\eqref{eqn:efield} at angular positions of sky sources using bivariate cubic spline interpolation, and formed the power beam by following Equation~\eqref{eqn:power_beam}. It is unclear how the instrument responds to emission near the horizon. \citet{Bassett2021} have investigated the effect of the horizon shape on extracting 21cm signals from observation, and found the signal extraction can be significantly improved when they include accurate information of the topological horizon such as nearby vegetation, buildings, and mountains. Modeling an accurate horizon can be crucial, but it is beyond the scope of this study and future work can explore more details. For simplicity, we applied a sharp cut-off to the observable sky and considered only sources above the horizon. For more accurate cubic spline interpolation, we allowed a 5-degree buffer below the horizon for the beam interpolation.

In constructing sky source models, we considered two types of foreground sources: galactic and extragalactic point sources, and diffuse synchrotron emission from our Galaxy. For the point source model, we adopted the GaLactic and Extragalactic All-sky MWA (GLEAM) survey which consists of GLEAM~I \citep{Hurley-Walker2017} and GLEAM~II \citep{Hurley-Walker2019}. The former covers about 25,000 square degrees of sky outside the galactic plane between -90$^\circ$ and 30$^\circ$ in declination and the latter covers compact sources in the galactic plane. The main lobe of our primary beam at 160--180~MHz has the FWHM of about 10$^\circ$ centered at the declination of -30.7$^\circ$, which means the field of view of the HERA main beam falls into the coverage of the GLEAM~I survey and the sky model is good enough to provide a representative set of point sources for our analysis. Point sources with spectral indices derived from single power-law fits provided by the GLEAM catalogs are used for the simulations, giving about 260,000 point sources. We also restored peeled bright sources noted in Table~2 of \citet{Hurley-Walker2017} and Fornax A which has two bright radio lobes \citep{Bernardi2013}.

For the diffuse sky model, the Global Sky Model (GSM) provided by \citet{Zheng2017} is used with a python package \texttt{PyGSM}. They removed 1\% of the highest peak of residual after fitting maps through their iterative Principal Component Analysis (PCA) algorithm, which helps remove some bright point sources in the diffuse map. In this study the brightness temperature of the diffuse sky map was generated in \texttt{HEALPix}, Hierarchical Equal Area isoLatitude Pixelization \citep{Gorski2005}, with \texttt{Nside} of 256 yielding 786,432 pixels and converted to flux density assuming the brightness temperature is constant in each pixel area. We then treated each pixel area as a point source and fed the sky model to our visibility calculations. For the frequencies we simulate and the size of the HERA array, we found that an \texttt{Nside} 256 \texttt{HEALPix} map resulted in enough resolutions for the proper power spectrum estimation, which is consistent with the results of \citet{Lanman2020}.

We chose the LST of 2.25 hours for our single LST simulation, a relatively foreground-free zone but also good for point source calibration for HERA \citep{Kern2020a}. In addition, by choosing the LST, the gap of the GLEAM catalog present at RA$\sim$8 hours has a small effect on our visibility calculation as it is outside the main lobe of the primary beam. A radio interferometer has a different response to the point sources and the diffuse source. To better understand the role of each sky model on the visibility, we consider the two sky models separately if necessary. Otherwise, we account for those sky models together as the combined sky model.

\subsection{Non-redundancy in Raw Visibility Measurements}
\label{sec:mad_vis}
When the feed positions are perturbed, redundant baselines are expected to produce different visibilities because of their different primary beam responses. This effect can be different depending on the configuration of the baseline (e.g., length and direction) and the type of the sky model (e.g., point sources or diffuse sources). One metric to evaluate the effect of the non-redundancy in the primary beams on visibilities is to measure a standard deviation of visibility measurements with the same baseline separation,
\begin{align}
{\rm \sigma}_{\rm red, \alpha} &= \sqrt{\frac{\sum_{\{i,j\}_\alpha}|V_{ij}(\nu) - \langle V_{ij}(\nu) \rangle_{\alpha}|^2}{N_{\rm rbl}}},
\label{eqn:std}
\end{align}
where $\alpha$ indicates an index of a redundant-baseline group, $N_{\rm rbl}$ is the number of baselines in the redundant-baseline group and $\langle V_{\rm ij}(\nu) \rangle_{\alpha}$ is the mean visibility for the redundant baselines. Here the visibility is raw data for which antenna gains and calibration are not applied. We employed $\sigma_{\rm red, \alpha}$ for auto- and cross-correlations to examine the non-redundancy in the raw visibility.

In general, the foreground spectrum is smooth, which means characteristic features in $\sigma_{\rm red}$ with frequency are mainly due to the features imprinted in the primary beams. According to Figure~\ref{fig:main_lobe_xz_plane} and \ref{fig:main_lobe_yz_plane}, while at some frequencies the beam width gets either narrower or wider as the feed moves upward, at around 170 MHz the beam width is nearly constant with the feed motion. Similar patterns are observed in $\sigma_{\rm red}$ of the visibility measurements with the axial feed motion as shown in the first panel of Figure~\ref{fig:std_auto}. The auto-correlation $\sigma_{\rm red}$ normalized by the averaged visibility amplitude over all antennas shows there is a null at around 170 MHz and the non-redundancy gets larger as it approaches the band edge. This value increases with the feed offset but the effect is small.

\begin{figure*}[t!]
\centering
\includegraphics[scale=0.46]{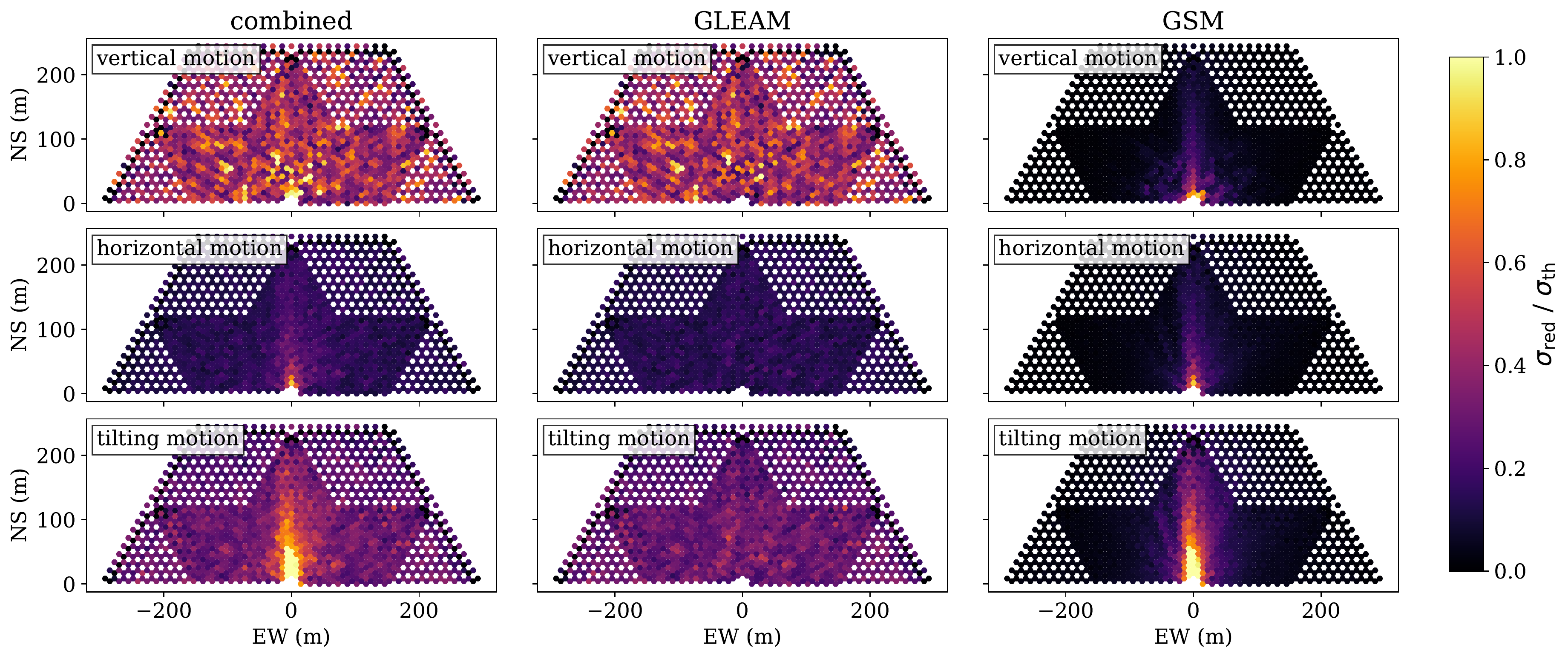}
\caption{Standard deviation of visibilities within a nominally redundant group of each baseline vector for the vertical (top row), horizontal (middle row), and tilting feed motion (bottom row). The first column is the result with the combined sky model which is separated into the GLEAM (second column) and GSM (third column) sky model. The color scale indicates $\sigma_{\rm red}$ normalized by the thermal noise of the visibility ($\sigma_{\rm th}$). The same value for $\sigma_{\rm th}$ which is 2.5 Jy from the fiducial model averaged over the frequency band is used for all panels. We consider visibilities with $\sigma_{\rm feed} = 3$ cm for the feed translation and $\sigma_{\rm feed} = 3^\circ$ for the tilting. We see diffuse emission introduces strong non-redundancy for particular baseline types, while point sources generate non-redundancy across all baselines.}
\label{fig:std_cross}
\end{figure*}

As we see in Figure~\ref{fig:main_lobe_xz_plane} and \ref{fig:main_lobe_yz_plane}, when the feed is displaced from the fiducial position in the $xy$ plane, the pointing angle of the main beam also shifts which is proportional to the feed displacement but frequency-independent. The middle panel of Figure~\ref{fig:std_auto} shows $\sigma_{\rm red}$ which is largely consistent with the result of the beam characteristics of lateral feed motions, showing larger variations in the auto-correlations with larger perturbation in the feed motion but nearly constant with frequency.

Similarly, as discussed in Section~\ref{sec:characteristics_beams}, pointing errors of beams due to the tilting motion is roughly linear to the tip-tilt of the feed. In addition, unlike the lateral feed motion, tilting produces non-negligible changes in the beam width. Combining the two effects results in a slightly different trend (right panel) compared to the middle panel.

The standard deviation computed from cross-correlations is more interesting and important in the sense that 1) responses of short and long baselines to point sources and the diffuse source are different and 2) cross-correlations are involved with redundant-baseline calibration and power spectrum estimation.

We derived $\sigma_{\rm red, \alpha}$ for each unique baseline using Equation~\eqref{eqn:std} which is then averaged over the frequency band. Figure~\ref{fig:std_cross} visualizes $\sigma_{\rm red}$ of cross-correlation in the baseline space with combined (left), GLEAM (middle), and GSM (right) sky models. From top to bottom, vertical, horizontal, and tilting motions are given. We chose $\sigma_{\rm feed} = 3$~cm for axial and lateral motions and $\sigma_{\rm feed} = 3^{\circ}$ for the tilting motion. Though the visibility simulation is noiseless, it is useful to compare $\sigma_{\rm red}$ with the expected thermal noise defined as,
\begin{align}
\sigma_{ij} = \sqrt{\frac{V_{ii}V_{jj}}{\Delta\nu \Delta t}},
\label{eqn:thermal_noise}
\end{align}
where $V_{ii}$ is the visibility autocorrelation of antenna $i$, $\Delta\nu$ = 0.125~MHz is the frequency channel width, and $\Delta$t~=~10~s is the integration time. The variations of auto-correlations due to feed motion are of order 1\% or less (Figure~\ref{fig:std_auto}), which means the thermal noise levels derived from Equation~\eqref{eqn:thermal_noise} are similar for all feed motions. For simplicity, we took a single value from the fiducial model that is $\sigma_{\rm th} = \langle\sigma_{ij}\rangle$~=~2.5~Jy averaged over the frequency band. $\sigma_{\rm red}$ for the GLEAM and the GSM is also divided by the same value for consistency.

The ratio of the non-redundancy in visibilities to the thermal noise is a key quantity to determine whether the redundant-baseline calibration works properly when there are beam perturbations. In most panels, the color scale less than 1 indicates the thermal noise is larger than the non-redundancy error and thus the effect of the non-redundancy on redundant-baseline calibration seems not significant. However, antenna gains which are solved for by the calibration come from at least two or more visibility measurements, making the noise levels in antenna gains smaller compared to those in visibilities and the non-redundancy effect on the calibrated data visible. Indeed, we show reduced $\chi^2$ from the redundant-baseline calibration including random thermal noises varies with the feed motion in Section~\ref{sec:redcal}.

In the first column of Figure~\ref{fig:std_cross}, the non-redundancy defined in the baseline space is different for different feed motion. While the vertical feed motion shows the evidence of non-redundancy across all baselines, the feature is concentrated at short EW baselines for the tilting motion. The horizontal motion presents similar patterns to those from the tilting motion but the overall amplitude is smaller. These different trends of $\sigma_{\rm red}$ can be thought as the results of different responses of deformed primary beams due to the feed motions to GLEAM and GSM sky models.

For the GLEAM-only sky model (middle column), $\sigma_{\rm red}$ is uniform across the baselines regardless of the feed motion. This is because a point source has a constant visibility amplitude for all baselines. One notable thing is $\sigma_{\rm red}$ is largest for the vertical motion and smallest for the horizontal motion, which is attribute to the different characteristics of the main lobe of the primary beam for different feed motions as discussed in Section~\ref{sec:characteristics_beams}. For example, when the feed moves along the vertical axis, the main beam widens/shrinks and collects more/less fluxes from the sky, which results in variation of the overall amplitude of visibilities and thus large $\sigma_{\rm red}$. In comparison, when the feed moves in the horizontal plane, the pointing angle of the main lobe shifts but with the nearly constant beam width. This shift of the beam makes the instrument observe a different part of sky but the total amount of received flux stays the same if the sky changes slowly in flux, which keeps $\sigma_{\rm red}$ small. The change of the main lobe width for the tilting motion is moderate and thus $\sigma_{\rm red}$ is between those of vertical and horizontal motions.

The diffuse model, however, behaves in a different way. Because the response of the interferometer to an extended source is strongest at the shortest baseline and declines with baseline length, $\sigma_{\rm red}$ is expected to be largest at short baselines as shown in the last column of Figure~\ref{fig:std_cross} for all three types of the feed motions. At LST~=~2.25~hours when the galactic plane is located at about 80$^\circ$ away from zenith, the overwhelming power from the galactic plane with large solid angles around the horizon makes the emission from diffuse sources near the horizon even stronger than that along zenith despite considerable attenuation of the primary beam outside the main lobe \citep{Thyagarajan2015a, Thyagarajan2015b}. This means, unlike the GLEAM case when $\sigma_{\rm red}$ is largely governed by the properties of main lobes, the features of side lobes play a role to characterize non-redundancy in visibilities arising from the diffuse sources. In this context, large $\sigma_{\rm red}$ in the tilting motion shown in the bottom right panel can be understood based on the large variation of $\epsilon_{\rm SL}$ defined in Equation~\eqref{eqn:side_lobe_int} (Figure~\ref{fig:side_lobe_int}).

Another interesting feature of the GSM-only model is the pattern in $\sigma_{\rm red}$ elongated along the NS direction which may originate from directional response of the interferometer. The geometric delay, $\tau_{\rm g}$, between two antennas forming a baseline is constant along the direction normal to the baseline vector on the sky. Therefore, the fringe defined by the geometric delay, $f \sim \exp{(-2\pi i \nu \tau_{\rm g})}$, has a constant value along the path orthogonal to the baseline vector. On the sky we are looking at, the galactic plane is stretched in the EW direction, which means the constant fringe in the EW direction for the NS baseline helps add up the flux from the galactic plane constructively and enhance the amplitudes of visibility measurements, showing the elongated pattern along the NS direction in $\sigma_{\rm red}$. Because the galactic plane moves around the sky as the Earth rotates, this effect is LST-dependent. For example, we found that at LST = 1~hour when the galactic center is located at the South-West corner of the sky aligned in the North-West direction, the pattern in $\sigma_{\rm red}$ is featured in the North-East direction as expected.

If we focus on the combined model again, now we can see the uniform trend of $\sigma_{\rm red}$ in the vertical feed motion is due to the strong response of the motion to the point source sky while large $\sigma_{\rm red}$ characterized at short EW baselines in the tilting motion is due to the strong response of the motion to the diffuse sky. Overall, the non-redundancy for the vertical and tilting motions is larger than that for the horizontal motion. \citet{Orosz2019} showed that intrinsic chromaticity on long baselines is in charge of introducing chromatic gain errors when non-redundancy is present in redundant-baseline calibration. HERA with the compact array layout has more short baselines than long baselines, which means the large non-redundancy featured at short baselines resulting from the GSM component may have considerable impact on the calibration as well.

\section{Redundant-baseline Calibration}
\label{sec:calibration}
In this section, we describe the antenna gains that are applied to the raw visibility measurements and need to be calibrated out in real observations.  We also describe the results of the calibration in the presence of non-redundancy in the raw visibility.

\subsection{Model of Input True Antenna Gains}
\begin{figure}[t!]
\centering
\includegraphics[scale=0.53]{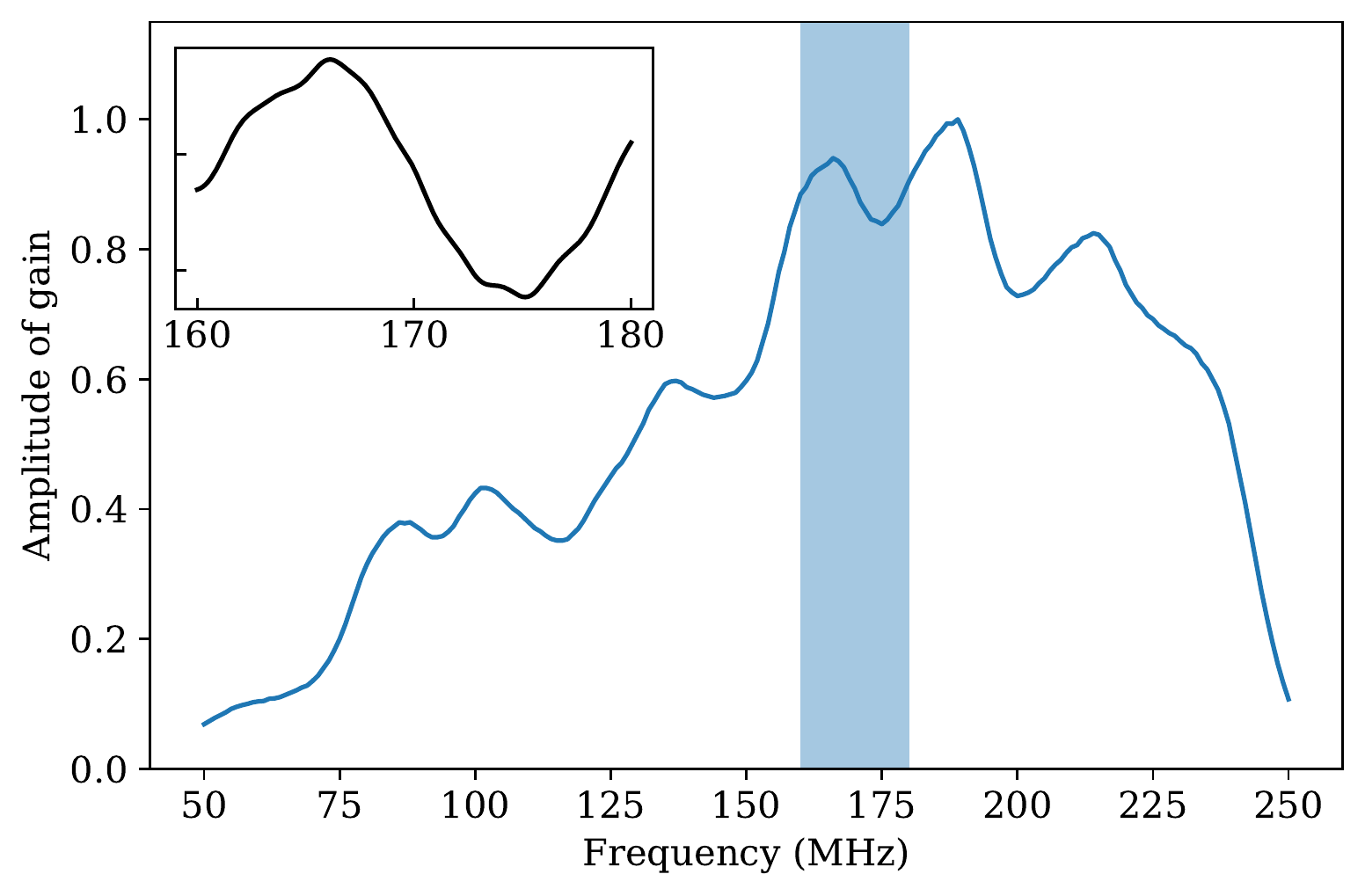}
\caption{The amplitude of the gain derived from Equation~\eqref{eqn:true_gain}. The response is normalized to the peak of the full band and has an arbitrary linear unit. The shaded region indicates the frequency band of 160--180~MHz which we use for visibility simulations. The inset panel shows the spectral features of the smoothed gain in the band.}
\label{fig:true_gain}
\end{figure}

The incident electric field on the feed and the output voltage are related by the system voltage response, $\textbf{H}$, defined as
\begin{align}
    V_{\rm out}(\nu, \theta, \phi) = \textbf{H}(\nu, \theta, \phi) \cdot \textbf{E}_{\rm in}(\nu, \theta, \phi).
\end{align}
The system, consisting of the antenna, FEM, coaxial cables, and post-ampliﬁer module, has $\textbf{H}$ expressed as
\begin{align}
    \textbf{H}(\nu, \theta, \phi) = \frac{Z_{\rm ant}(\nu) + Z_0}{\nu Z_0} \textbf{E}_{\rm pat}(\nu, \theta, \phi) S_{21}(\nu),
    \label{eqn:true_gain}
\end{align}
where $Z_{\rm ant}$ is the complex impedance of the antenna, $Z_0 = 100 \, \Omega$ is a termination impedance which is appropriate for a differential signal, $\textbf{E}_{\rm pat}$ is the electric-field pattern, and $S_{21}$ is a scattering parameter \citep{Fagnoni2020}. The $S_{21}$ scattering parameter describing the power transferred from port~1 to port~2 in the receiver comes from lab measurements, and $Z_{\rm ant}$ and $\textbf{E}_{\rm pat}$ are obtained from CST simulations. This $\textbf{H}$ represents the antenna gain relating sky signals to instrument measurements via the antenna and RF chain. Because we take into account a direction-independent gain in the following analysis, we picked the system voltage response at boresight where the beam response is strongest, $\textbf{H}(\nu, 0, 0)$. The antenna gain is normalized to the peak for the full band and the result is presented in Figure~\ref{fig:true_gain} with the highlighted frequency band of interest. Detailed derivation of the antenna gain is addressed in \citet{Fagnoni2020} and Hewitt et al. (HERA memo \#105\textsuperscript{\ref{footnote1}}).

The $S_{21}$ measurements were sampled at 1~MHz cadence and thus must be interpolated to the finer resolution of 0.125~MHz. To prevent unwanted fine-scale spectral artifacts from the interpolation, the gain amplitudes at 160--180 MHz were smoothed and interpolated by a Gaussian process regression (GPR) model with a fixed kernel size of 1~MHz after subtracting a 5th-order polynomial fit. The GPR may smooth out potential frequency structure smaller than 1~MHz but is found to be effective in suppressing interpolation artifacts. More details about the smoothing method are described in \citet{Lanman2020}. The result of the smoothed gain is shown in the inset panel of Figure~\ref{fig:true_gain}. Though the $Z_{\rm ant}(\nu)$, $\textbf{E}_{\rm pat}$, and $S_{21}$ vary with the feed displacement, we ignore this effect and model the same gain from the unperturbed case, regardless of the feed motion for simplicity.

\begin{figure*}[t!]
\centering
\includegraphics[scale=0.5]{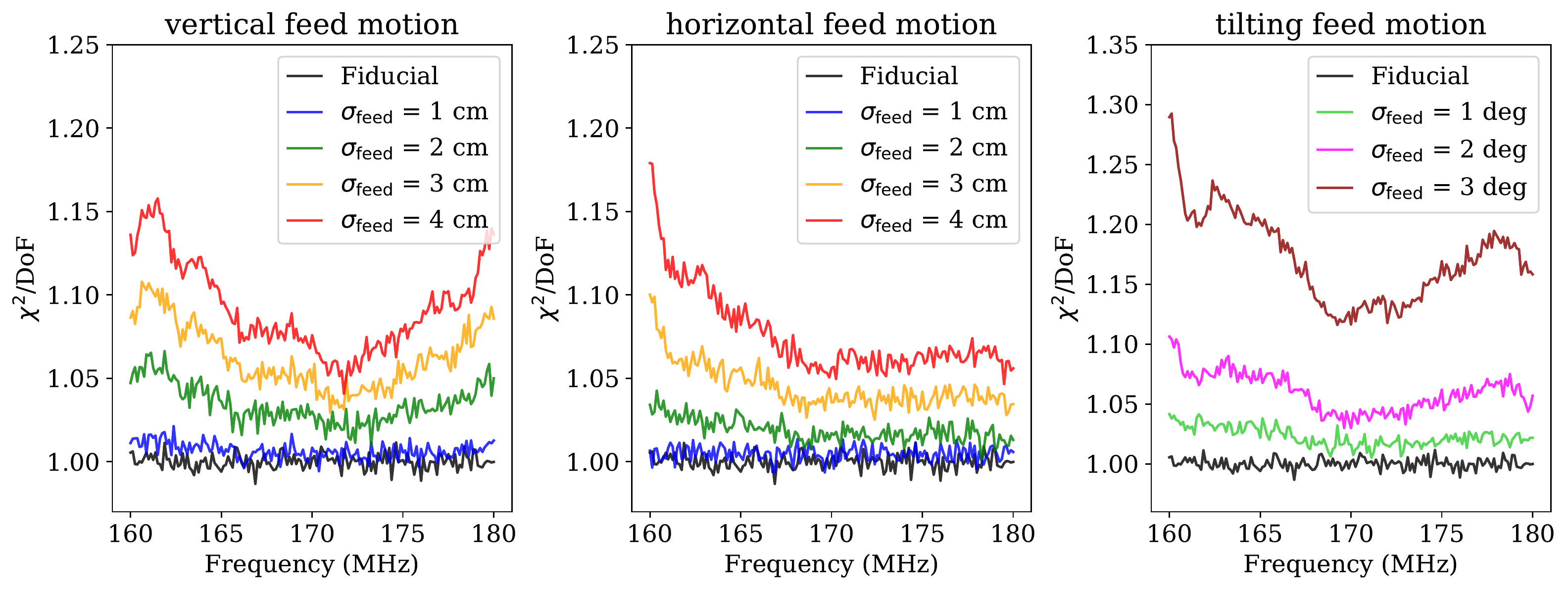}
\caption{$\chi^2$/DoF of the redundant-baseline calibration for simulated visibilities with perturbed primary beams. From left to right, vertical, horizontal, and tilting feed motions are presented. The fiducial model yields $\chi^2$/DoF close to 1 as expected, which indicates a perfect calibration. When the feed is displaced from the fiducial position, $\chi^2$/DoF also deviates from 1. This effect is largest for the tilting motion, which is consistent with $\sigma_{\rm red}$ of raw visibilities shown in Figure~\ref{fig:std_cross}. The noisy features are due to the thermal noise that is included only for the $\chi^2$ statistics.}
\label{fig:chi2}
\end{figure*}

To account for the realistic effect of the amplitude attenuator and the cable delay, we randomized amplitudes and phases of antenna gains over 320 antennas. We assume the attenuation $\eta_i$ for each antenna follows a random Gaussian distribution with zero mean and a standard deviation of 0.2 and the cable delay $\tau_i$ also follows a random Gaussian distribution with zero mean and a standard deviation of 20~ns corresponding to 6-m long cable. The smoothed gain was then multiplied by $\exp(\eta_i + 2\pi i \tau_i \nu)$ and the final ``true'' gains were applied to the raw visibility to derive the measured visibility.

\subsection{Effects of Non-redundancy on Redundant-baseline Calibration}
\label{sec:redcal}
The general relation between the measured visibility $V_{ij}^{\rm obs}$ and the true visibility $V_{ij}^{\rm true}$ is assumed to be associated with antenna gains and thermal noise,
\begin{align}
    V_{ij}^{\rm obs} = g_i g_j^*V_{ij}^{\rm true} + n_{ij},
    \label{eqn:calib_eqn}
\end{align}
where $g_i$ and $g_j$ are complex per-antenna gains and $n_{ij}$ is the Gaussian thermal noise. The antenna gain represents direction-independent effects such as amplifiers and the delay offset due to the light-travel time delay along the path. The antenna-to-antenna variation of primary beams, which is our interest, is one common source of the direction-dependent effect \citep{Smirnov2011}. Equation~\eqref{eqn:calib_eqn}, however, does not account for the direction-dependent correction, which may leave chromatic errors in the antenna gains. To minimize the potential spectral structure from the direction-independent calibration, some techniques such as smoothing the antenna gains can be employed. More details about mitigation with the techniques will be addressed in Kim et al. (in prep).

With all identical primary beam models, antenna pairs with the same separation are supposed to measure the same true visibility. The redundant-baseline calibration uses the redundancy of the visibility in the redundant baseline group rather than prior information of the sky to obtain the true visibility. More specifically, the redundant-baseline calibration solves for the unique visibility along with the antenna gains as free parameters,
\begin{align}
    V_{ij}^{\rm obs} = g_i g_j^*V_{i-j}^{\rm sol},
    \label{eqn:redcal_eqn}
\end{align}
where $V_{i-j}^{\rm sol}$ is the unique visibility solution of the redundant baseline group with the same baseline separation. One approach to find the antenna gain and the visibility solution is to minimize $\chi^2$,
\begin{align}
    \chi^2 = \sum_{i<j} \frac{|V_{ij}^{\rm obs} - g_i g_j^* V_{i-j}^{\rm sol}|^2}{\sigma_{ij}^2},
    \label{eqn:chi2}
\end{align}
where $\sigma_{ij}^2$ is the variance of $n_{ij}$. This is performed per polarization, per frequency, and per time. More comprehensive discussion about how to minimize Equation~\eqref{eqn:chi2} is presented in \citet{Dillon2020}. The calibration was performed with  the publicly available software library, \texttt{hera\_cal}\footnote{\url{https://github.com/HERA-Team/hera_cal}}.

\begin{figure*}[t!]
\centering
\includegraphics[scale=0.5]{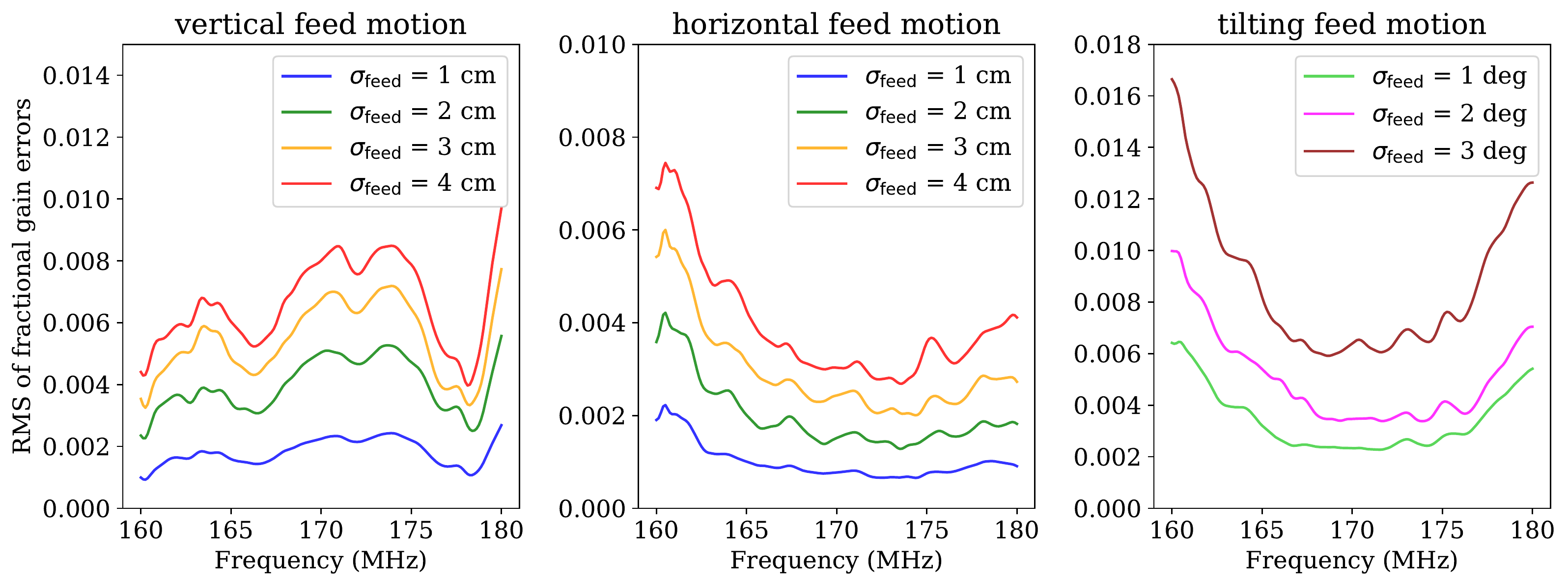}
\caption{RMS of fractional gain errors for the vertical (left), horizontal (middle), and tilting (right) feed motions. As expected, larger feed motion perturbations result in larger RMS of fractional gain errors. Although these gain errors are kept largely about 1\%, when convolved with bright foregrounds this could easily swamp the intrinsic 21-cm signal if the gain errors have sufficient structure.}
\label{fig:gain_err}
\end{figure*}

In an ideal case when redundant baselines share the same true visibility, $\chi^2$ is expected to be equal to the degree of freedom (DoF). The DoF for the redundant-baseline calibration is ${\rm DoF} = N_{\rm bl} - N_{\rm ubl} - N_{\rm ant} + 2$ where $N_{\rm bl}$ is the total number of baselines, $N_{\rm ubl}$ is the number of unique baselines, and $N_{\rm ant}$ is the number of antennas \citep{Dillon2020}. Figure~\ref{fig:chi2} shows overall $\chi^2$/DoF as a function of frequency for vertical (left), horizontal (middle) and tilting (right) feed motions\footnote{Even though our simulations are noise-free throughout the rest of our analysis, for appropriate $\chi^2$ calculation, visibility measurements to derive Figure~\ref{fig:chi2} include thermal noise generated with autocorrelations using Equation~\eqref{eqn:thermal_noise}. We consider $\Delta\nu$~=~125~kHz and $\Delta$t~=~10~seconds.}. The measured visibility is simulated with the combined sky model. As expected, the fiducial model yields $\chi^2$/DoF close to 1, which indicates a perfect calibration. When the feeds move away from the fiducial position and angle, $\chi^2$/DoF deviates from 1 and the deviation gets larger with the degree of perturbation. For a given $\sigma_{\rm feed}$, vertical feed motions result in a similar level of $\chi^2$/DoF to horizontal feed motions, but the former has a slightly larger mean value. Our choice of tilts, however, presents rather larger $\chi^2$/DoF compared to the translational motions. This is largely consistent with the results of $\sigma_{\rm red}$ for raw visibilities, showing the larger non-redundancy error arising from the tilting motion compared to that from the horizontal motion (Figure~\ref{fig:std_cross}).

\begin{figure*}[t!]
\centering
\includegraphics[scale=0.5]{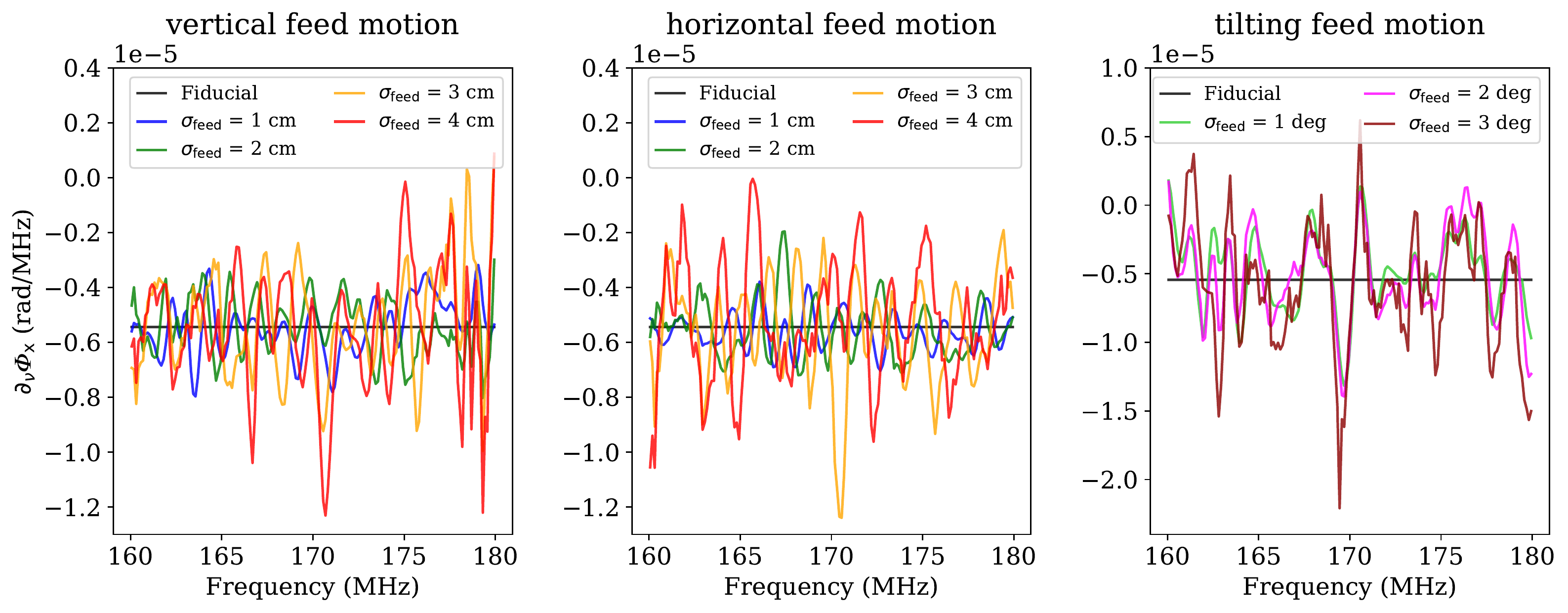}
\caption{Slope of the $x$-direction phase gradient that is solved for by absolute calibration for vertical (left), horizontal (middle), and tilting (right) feed motions. $\partial_\nu \Phi_{\rm x}$ represents the derivative of the linear phase gradient in the $x$-direction with respect to frequency. We see the slope is nearly constant for the fiducial model, while there is high-frequency structure for the perturbed model which introduces additional chromatic gain errors.}
\label{fig:degen_param}
\end{figure*}

Another metric to quantify the effectiveness of the redundant-baseline calibration against perturbed beams with feed motions is the RMS fractional gain errors. The fractional gain error of an antenna is defined as
\begin{align}
f_{g, i} = \frac{|g_{{\rm pert}, i}| - |g_{{\rm unpert}, i}|}{|g_{{\rm unpert}, i}|}.
\label{eqn:frac_gain_err}
\end{align}
Here $|g_{{\rm unpert}, i}|$ and $|g_{{\rm pert}, i}|$ indicate the gain amplitudes of the unperturbed and perturbed beams for antenna $i$, respectively. The RMS of fractional gain errors is then calculated over all 320 antennas per frequency. In Figure~\ref{fig:gain_err}, the vertical (left) and horizontal (middle) feed motions show $\lesssim~1\%$ RMS errors, while the tilting motion (right) displays larger errors, about 1--2 \% RMS errors, which is consistent with the results of $\chi^2$/DoF. In other words, larger chromatic gain errors indeed correlate to larger feed motion perturbations.

Though the highly redundant array configuration of HERA yields enough number of measurements to solve for the unknown parameters of redundant-baseline calibration, there are degeneracies between the antenna gains and the visibility solution that keep $g_i g_j^* V_{i-j}^{\rm sol}$ unchanged. The four degenerate parameters are the overall amplitude and phase, along with the EW and NS tip-tilt \citep{Liu2010, Zheng2014, Dillon2018, Li2018, Byrne2019, Kern2020a}. The last two are due to directional phase factors in $x$- and $y$-direction in the antenna gain which can be cancelled by rephasing the unique visibility solution (e.g., $g_i \rightarrow g_i e^{i\Phi_x x_i}$ and $V_{i-j}^{\rm sol} \rightarrow V_{i-j}^{\rm sol}e^{-i\Phi_x (x_i-x_j)}$, where $\Phi_x$ is a linear phase gradient in the $x$-direction). The overall phase is the degeneracy between antenna gains ($g_i \rightarrow e^{i\psi} g_i$, $g_j^* \rightarrow e^{-i\psi} g_j^*$) and is set arbitrarily. The three degeneracies other than the overall phase are solved for by an additional process referred as absolute calibration using a sky model.

We implemented absolute calibration with the model visibility simulated using the unperturbed primary beam model. As described in Section~\ref{sec:characteristics_beams}, the pointing angle of the primary beam can shift due to feed motion, resulting in shifted field of view of the observed sky relative to the desired fiducial pointing. Sky-based calibration with an inaccurate sky model is known to introduce frequency-dependent calibration errors \citep{Barry2016, Ewall-wice2017, Byrne2019, Gehlot2021}. In other words, the difference in the sky observed by unperturbed and perturbed primary beams is an additional potential source of chromatic gain errors \citep{Orosz2019, Barry2022}. Figure~\ref{fig:degen_param} is an example showing the linear phase gradient along the $x$-direction that is solved for by the absolute calibration step. We calculate the derivative of the phase gradient with respect to frequency, and present the results as a function of frequency for the vertical (left), horizontal (middle), and tilting (right) motions. The derivative for the fiducial model is nearly constant. However, the phase gradient for the perturbed case exhibits high-frequency structure, which contributes to the spectral leakage especially at high delay modes.

\begin{figure*}[t!]
\centering
\includegraphics[scale=0.55]{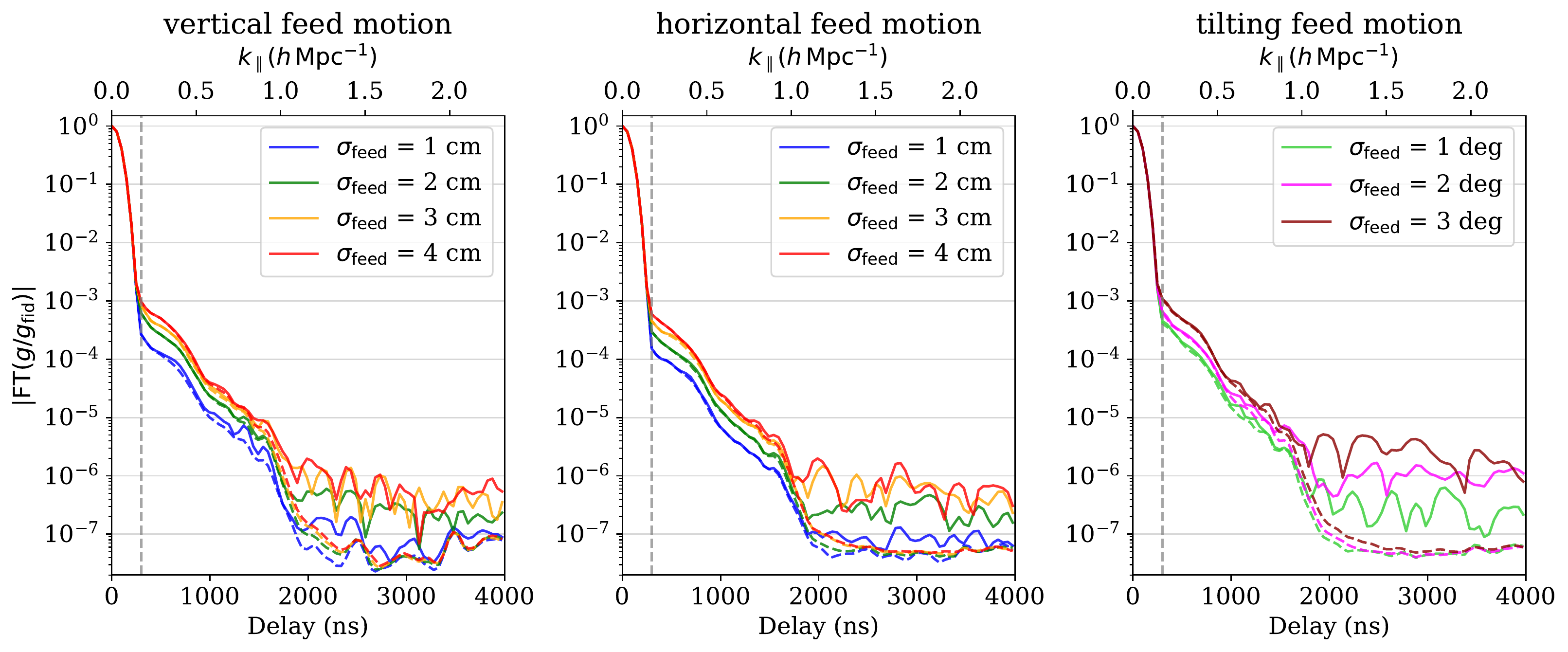}
\caption{Averaged Fourier transformed antenna gain over all antennas for the axial (left), the lateral (middle), and the tilting (right) feed motions. The perturbed gain is normalized to the fiducial one. The dashed and solid lines indicate the results after the redundant-baseline and after the full calibration including the absolute calibration, respectively. Different colored lines denote different levels of perturbation in feed motions. Even in the case when the feed is least perturbed in our choice, a broad wing $\gtrsim$ 300 ns or 0.17 $\hMpcinv$ (vertical line) is present. This is due to chromatic gain errors caused by the non-redundancy in visibilities emerging from redundant-baseline calibration. Existence of the wing is critical since it affects detection of cosmological signals at low $\kpara$ where high signal-to-ratios are expected. The chromatic gain errors above 2000~ns mainly come from the absolute calibration.}
\label{fig:fft_gain}
\end{figure*}

Figure~\ref{fig:fft_gain} demonstrates a frequency Fourier transform of the calibrated antenna gains after the redundant calibration (dashed lines) and the full calibration including the post absolute calibration (solid lines). The Fourier transform of the gain shows the frequency structure in the gain that will be mapped to power spectrum estimates. The delay in the $x$-axis is a Fourier dual to frequency. To derive the $y$-axis, we divide the perturbed gains by the fiducial one, and multiply them by a 7th-order Blackman-Harris tapering window function which helps suppress side lobes and achieves a large dynamic range \citep{Lanman2020}. We then perform the frequency Fourier transform, normalize the results to the peaks, and average the amplitudes over all antennas.

We found that the Fourier transform of the fiducial gain, with the perfect calibration, reaches a floor of about $10^{-9}$ at high delays, which is consistent with the true gain. This means the floor above about $10^{-7}$ shown in Figure~\ref{fig:fft_gain} may arise from the calibration error due to the perturbed beams. In addition, unwanted broad wings start to appear at delay larger than 300~ns, which are primarily caused by chromatic errors in the gain solutions from the redundant-baseline calibration as shown in dashed lines. Larger perturbation in feed motion results in larger non-redundancy errors, and thus larger amplitude of the wing. In principle, the antenna gain from the perfect calibration that is convolved with smooth spectrum of foregrounds makes the foreground power isolated in the wedge of a power spectrum. When the gain includes chromatic errors, however, convolution of the gain and the foregrounds can lead to foreground leakage outside the wedge owing to the broad wing in the Fourier transform. Additional chromatic gain errors at delay larger than $\sim$2000~ns are observed mainly due to the absolute calibration step.

\section{Power Spectrum Estimation with Feed Motion}
\label{sec:pspec}
An infinite frequency Fourier transform of a visibility for spectrally flat foreground emission from a single source forms a Dirac delta function in the delay domain. Realistic consideration with smoothly varying foreground and instrumental responses with frequency as well as the finite band width turns the Dirac delta function into a broadened delay-spectrum, but the width remains small as long as the variation is smooth enough. The delay-spectrum is centered at the geometric delay of the source, $\tau_{\rm g} = |\textbf{b}|\sin{\theta}/c$ where $|\textbf{b}|$ is the norm of the baseline vector, $\theta$ is the angle from zenith to the source, and $c$ is the speed of light. The frequency Fourier transform of a visibility for all foreground sources is then the superposition of such delay-spectra across the field-of-view of the sky, and the maximum delay mode is defined by the largest geometric delay between two antennas forming the visibility. With a zenith-pointed array, the maximum geometric delay is set by the horizon limit and thus the delay modes are bound to $\tau \lesssim \tau_{\rm hor} = |\textbf{b}|/c$. The complex-spectrum cosmological signal due to line-of-sight fluctuations, however, is widely distributed across all delay modes even beyond the horizon limit, which leaves a room for detecting the cosmological signals at $\tau > \tau_{\rm hor}$.

\subsection{The Power Spectrum for Cosmological Signals}
The observable cosmological quantity measured from redshifted 21-cm observation is the brighteness temperature of the neutral hydrogen that is defined by the spin temperature $T_{\rm S}$ relative to the background radiation temperature $T_\gamma$ \citep[e.g.,][]{Furlanetto2006},
\begin{align}
\delta T_{\rm b}(z) &= \frac{T_{\rm S} - T_{\gamma}(z)}{1+z}(1-e^{-\tau_{\nu_{0}}}) \\
&\approx 27 x_{\rm HI} (1+\delta_{\rm m}) \bigg(\frac{H(z)}{{\rm d}v_\parallel/{\rm d}r_\parallel + H(z)}\bigg) \bigg(1 - \frac{T_\gamma}{T_{\rm S}}\bigg) \nonumber \\
&\,\times \bigg(\frac{1+z}{10}\frac{0.15}{\Omega_{\rm m} h^2}\bigg)^{\frac{1}{2}} \bigg(\frac{\Omega_{\rm b} h^2}{0.023}\bigg)  \quad  [{\rm mK}],
\label{eqn:21cm_brightness_temp}
\end{align}
where $\tau_{\nu_{0}}$ is the optical depth at the rest 21-cm frequency $\nu_{0} = 1420$~MHz, $x_{\rm H\textsc{I}}$ is the fraction of the neutral hydrogen, $\delta_{\rm m}$ is the matter density fluctuation, $H(z)$ is the Hubble parameter, and ${\rm d}v_\parallel/{\rm d}r_\parallel$ is the gradient of the line-of-sight velocity.

We adopt a cosmological signal from \citet{Mesinger2016} generated with \texttt{21cmFAST} \citep{Mesinger2011} using the Faint Galaxies model that is consistent with Ly~$\alpha$ forest observations. The size of the simulation box is $1600^3$~cMpc$^3$ with 1024 pixels on each side. We chose the coeval cube at $z\sim7.4$ corresponding to 170~MHz. The cosmological model at this redshift predicts the ionization fraction of hydrogen gas $\lesssim$ 0.5.

\begin{figure}[t!]
\centering
\includegraphics[scale=0.6]{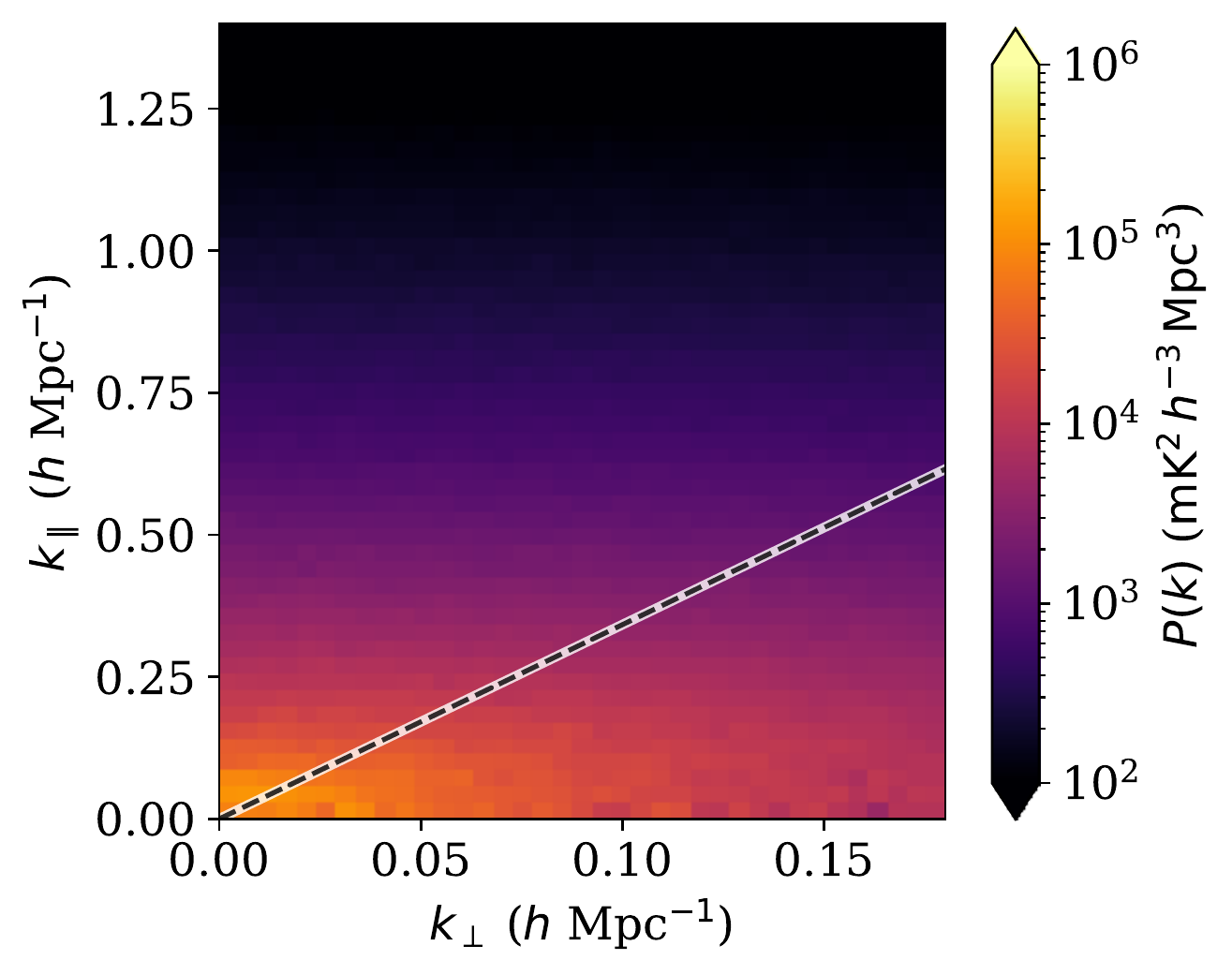}
\caption{EoR power spectrum estimated at $z\sim7.4$. Spectral structure in the cosmological signal allow the power spectrum to spread beyond the horizon limit (black dashed line) defined in Equation~\eqref{eqn:horizon_limit}, forming the isotropic power spectrum.}
\label{fig:pspec_eor}
\end{figure}

Figure~\ref{fig:pspec_eor} shows the power spectrum of EoR signals defined in ($\kperp$, $\kpara$) space where $\kperp$ and $\kpara$ are Fourier modes perpendicular to and parallel to the line-of-sight, respectively. The cosmological power spectrum is estimated with
\begin{align}
    \hat{P}_{21}(\textbf{k}, z) = \frac{\langle|\delta\Tilde{T}_{\rm b}(\textbf{k}, z)|^2\rangle}{V}  \quad  [{\rm mK^2} \,  h^{-3} \, {\rm Mpc^{3}}],
\end{align}
where $\delta \Tilde{T}_{\rm b}(\textbf{k}, z)$ is a three-dimensional spatial Fourier transform of $\delta T_{\rm b}(\textbf{x}, z)$ and $V$ is the simulation volume. The effective frequency bandwidth with the 7th-order Blackman-Harris tapering function is about 5.4~MHz or $\Delta z = 0.27$ in which the universe can be regarded as coeval and the light-cone effect on the power spectrum can be ignored \citep{Datta2012}. As expected, the EoR signal forms an isotropic power in the two-dimensional power spectrum that is extended to high $\kpara$, providing a region for detecting the EoR above the horizon limit. This power spectrum will be used as a reference to quantify the foreground leakage beyond the horizon limit.

\subsection{The Foreground Leakage due to Non-uniform Primary Beam Models}

In principle, the foregrounds and the cosmological signal are separable in the two-dimensional power spectrum thanks to the different behaviors of the sources in the Fourier domain. The power spectrum can be obtained based on the Fourier transform of a visibility along frequency,
\begin{align}
    \Tilde{V}(\textbf{u}, \tau) = \int w(\nu) V(\textbf{u}, \nu) e^{2\pi i \nu \tau} d\nu.
\end{align}
Here, $\textbf{u} = \textbf{b}/\lambda$ and $w(\nu)$ is a tapering function applied along the frequency axis to down-weight the edge effect of the bandpass. As described in Section~\ref{sec:redcal}, we chose the 7th-order Blackman-Harris window function.

\begin{figure}[t!]
\centering
\includegraphics[scale=0.6]{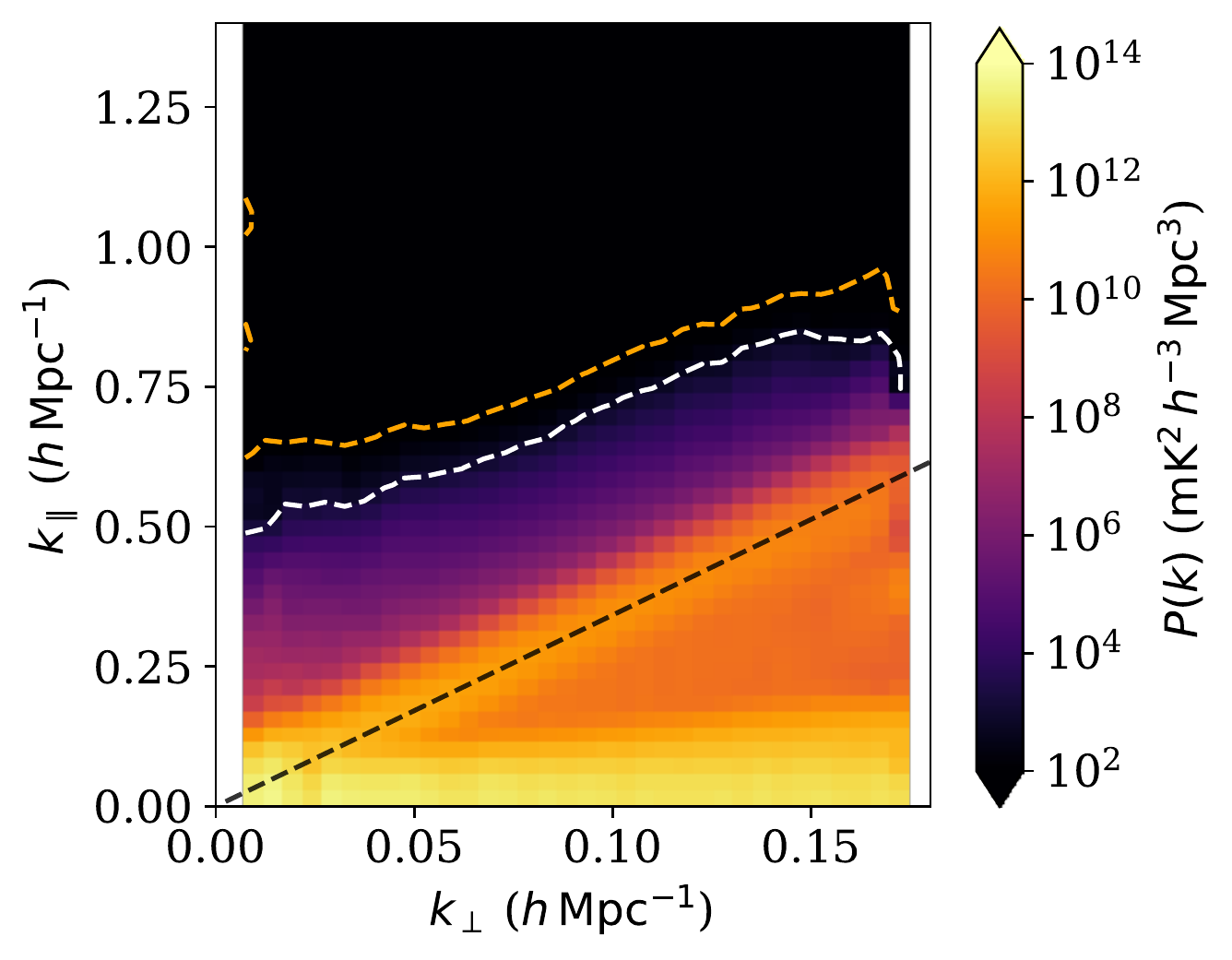}
\caption{Foreground power spectrum for the fiducial beam model with the combined sky model. The power spectrum is estimated with the calibrated visibility at LST = 2.25~hours for the 160--180~MHz band. The black dashed line indicates the geometric horizon limit. The strong emission at low $\kpara$ is due to the point sources located at zenith while the emission aligned with the horizon limit is due to the diffuse source near the horizon. The white and orange contours are where the foreground spillovers are the same as and 10\% of the EoR power, respectively. The foreground spillover beyond the horizon limit is not due to calibration errors but due to the intrinsic chromaticity of the beam in the far side lobes convolved with the diffuse power originating from the pitchfork effect. Without the fringe-rate filter, the cosmological information below $k_{\parallel} \sim 0.5 \, \hMpcinv$ is obscured by the foreground power, even with the perfect calibration.}
\label{fig:pspec_fiducial}
\end{figure}

\begin{figure*}[t!]
\centering
\includegraphics[scale=0.48]{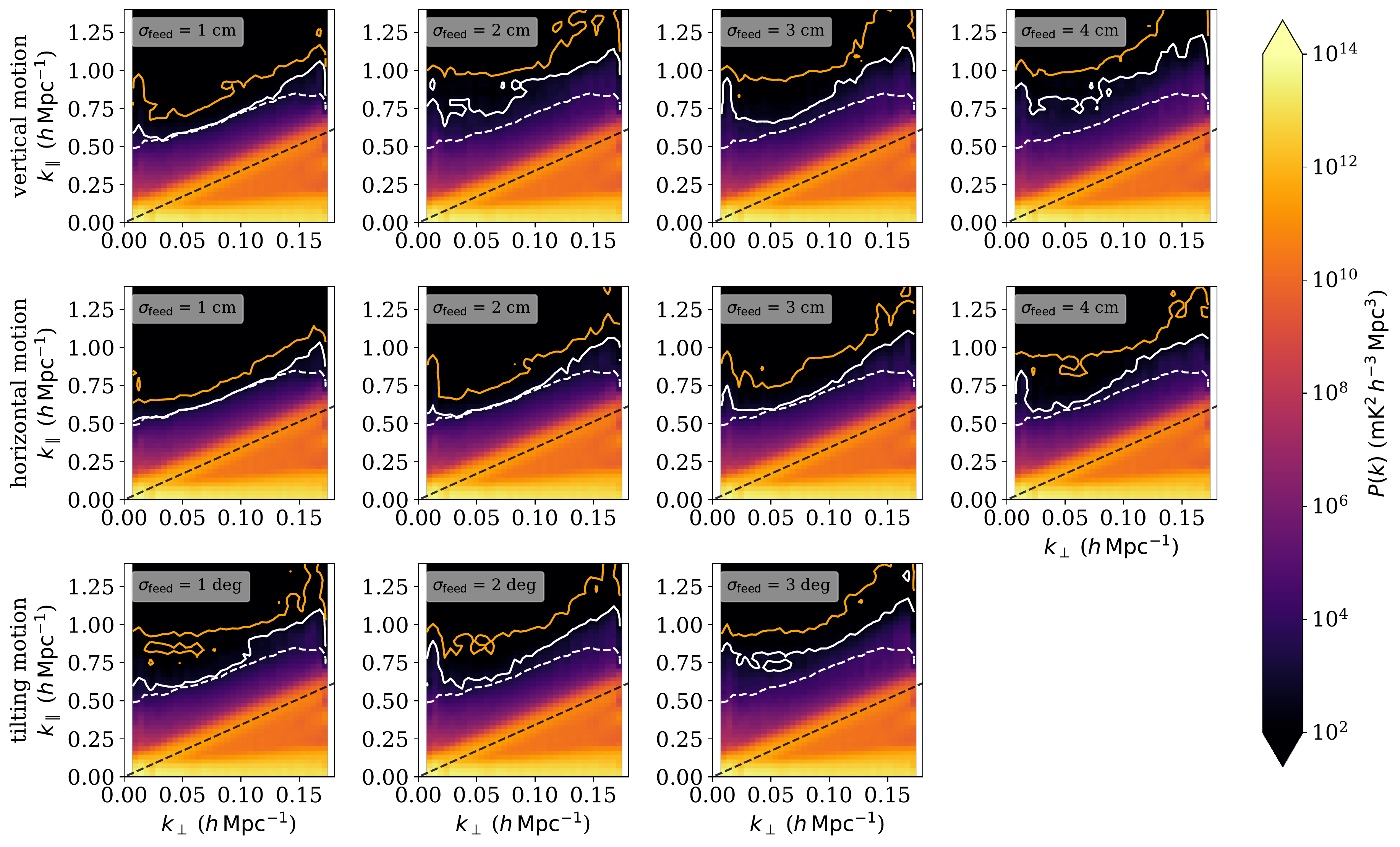}
\caption{Foreground power spectra for vertical (top row), horizontal (middle row), and tilting (bottom row) feed motions with the combined sky model. From left, power spectra of $\sigma_{\rm feed}$ = 1, 2, 3, and 4 cm for the translation motions and $\sigma_{\rm feed}$ = 1$^\circ$, 2$^\circ$, and 3$^\circ$ for the tilting motion are presented. The white and orange solid contours indicate where the foreground is equal to and 10\% of the EoR power, respectively. The white dashed contour representing the fiducial model is shown in all panels for comparison. As expected, there is contamination when the feed positions are perturbed. We do not apply any mitigation technique to reduce the chromatic gain errors such as down-weighting the effect of long baselines in calibration or smoothing the gain solutions along the frequency axis, which will be discussed in the subsequent paper.}
\label{fig:pspec_all_no_mitigation}
\end{figure*}

We coherently average complex visibilities with the same redundant baselines, which helps reduce the spectral structure arising from the chromatic gain errors. The coherently averaged visibilites, $V_{\rm coh}$, are converted to delay spectra and the square of them yields the power spectrum estimate,
\begin{align}
    \hat{P}(\kperp, \kpara) = \frac{X^2Y}{B_{\rm pp}\Omega_{\rm pp}} |\Tilde{V}_{\rm coh}(\textbf{u}, \tau)|^2,
    \label{eqn:pspec}
\end{align}
where $B_{\rm pp}$ is the effective band width defined as $B_{\rm pp} = \int |w(\nu)|^2 d\nu$ and $\Omega_{\rm pp}$ is the spatial integral of the squared primary beam \citep{Parsons2014}. $X$ and $Y$ are scaling factors relating $|\textbf{u}|$ and $\tau$ to $\kperp$ and $\kpara$ in cosmological units,
\begin{align}
    \kperp &= \frac{2\pi|\textbf{u}|}{X} \\
    \kpara &= \frac{2\pi\tau}{Y},
\end{align}
where $X = D(z)$, $Y = c(1+z)^2/(\nu_{0}H(z))$, and $D(z)$ is the comoving distance. The horizon limit, $\tau_{\rm hor} = |\textbf{b}|/c$, is then expressed in cosmological units as
\begin{align}
    k_{\rm \parallel, hor} &= \frac{H(z)D(z)}{c(1+z)} \kperp.
    \label{eqn:horizon_limit}
\end{align}
We utilized \texttt{hera\_pspec}\footnote{\url{https://github.com/HERA-Team/hera_pspec}} that is a publicly available software to construct the power spectrum estimation.

Figure~\ref{fig:pspec_fiducial} presents the estimated power spectrum of the combined sky model for the fiducial feed position. For the fiducial model, the foreground power is expected to fall off rapidly with $\kpara$ beyond the horizon limit (black dashed line). However, significant amount of foreground spillovers are observed beyond the horizon limit despite the perfect calibration. This is an intrinsic spillover associated with the powerful emission near the horizon known as the ``pitchfork effect'' along the black dashed line \citep{Thyagarajan2015a, Thyagarajan2015b}. The pitchfork effect can be understood as a result of the response of short projected baselines to strong diffuse sky emissions from the horizon lined up with the horizon limit in the ($\kperp$, $\kpara$) space.

\begin{figure*}[t!]
\centering
\includegraphics[scale=0.48]{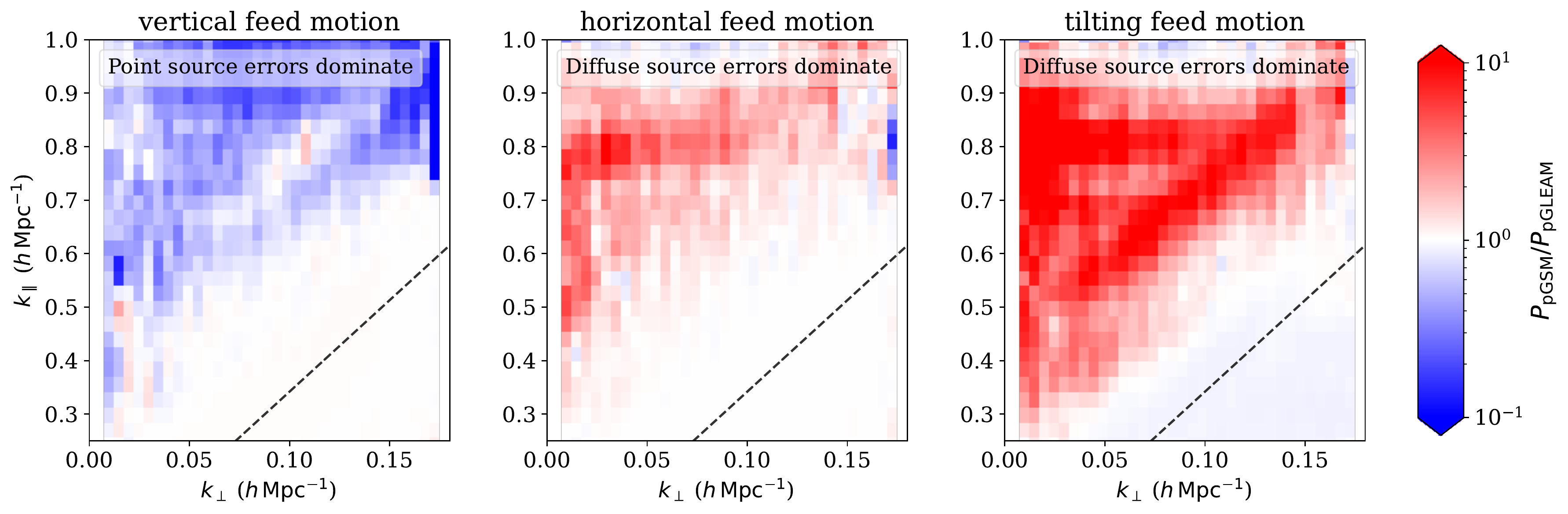}
\caption{Relative foreground power spectra, $P_{\rm pGSM}/P_{\rm pGLEAM}$, for vertical (left), horizontal (middle), and tilting (right) feed motions. $P_{\rm pGSM}$ is from the visibility that is simulated using the GSM with the perturbed beam and the GLEAM with the fiducial beam. $P_{\rm pGLEAM}$ indicates the opposite case when the GLEAM is with the perturbed beam and the GSM is with the fiducial beam. The black dashed line indicates the horizon limit. Base on the analyses with $\sigma_{\rm red}$ in Figure~\ref{fig:std_cross}, $P_{\rm pGLEAM}$ is expected to have larger foreground leakage than $P_{\rm pGSM}$ for the vertical motion while $P_{\rm pGSM}$ is in charge of leaking the foreground into the EoR window for the horizontal and tilting motions.}
\label{fig:pspec_perturb_one_source}
\end{figure*}

At LST $\sim$ 2~hours which is relatively foreground free in the direction of the zenith-pointed sky, the diffuse galactic plane lies near the horizon and forms the pitchfork. Because the galactic plane emanates strong synchrotron emissions, convolving the primary beam response with the sky emission in the Fourier domain results in an excess of power above the horizon limit as shown in Figure~\ref{fig:pspec_fiducial}. There is a trend that low $\kperp$ corresponding to short baselines has more excess of power than high $\kperp$ and this is because short baselines less resolve out the diffuse source and thus capture more power on the horizon leading to more power at high $\kpara$ than long baselines. The overplotted white and orange contours correspond to the power level where the foreground power is equal to and 10\% of the cosmological signal, respectively. These contours reveal the foreground spillovers extended above the horizon limit clearly.

This result is somewhat different from the result of \citet{Orosz2019} who showed smaller spillovers beyond the horizon limit. One major difference between their and our simulations is that they included point sources only while we consider diffuse sources as well as point sources. We found if we only consider point sources throughout the analysis, we came into a similar result to \citet{Orosz2019} with considerably suppressed spillover beyond the horizon limit. We include the GSM because it represents a more realistic sky model. The pitchfork effect can be mitigated by filtering out small fringe-rate regions on the sky \citep{Parsons2016}, and the effects of applying this technique will be addressed in Kim et al. (in prep).

In Figure~\ref{fig:pspec_all_no_mitigation}, from left to right, power spectra perturbed by $\sigma_{\rm feed}$ = 1, 2, 3, and 4 cm for the vertical (top row) and the horizontal (middle row) feed motion and $\sigma_{\rm feed}$ = 1$^\circ$, 2$^\circ$, and 3$^\circ$ for the tilting motion (bottom row) are shown. Across all feed motions, as the feed moves away from the fiducial point, the solid contours representing the perturbed models depart from the dashed contour which is the fiducial one shown in Figure~\ref{fig:pspec_fiducial}. This lift of the foreground leakage is associated with the antenna gains corrupted by the chromatic errors as discussed in Figure~\ref{fig:fft_gain}. At short baselines (i.e., low $\kperp$) which have a number of redundant baselines, coherent averaging in the visibility space can reduce high-frequency structure introduced by the chromatic gain errors. As a result, for $\sigma_{\rm feed}$ = 1 cm or $\sigma_{\rm feed}$ = 1$^\circ$ when the feed displacement is relatively small, the averaged visibility yields a similar foreground power to that of the fiducial model at low $\kperp$. Longer baselines, however, have relatively smaller number of redundant baselines and the averaging is not sufficient to smooth out the high-frequency features imprinted by the chromatic gain errors on top of the intrinsic spectral structure. This makes the long baselines suffer from unsmoothed frequency structure, displaying the leakage along $\kpara$ in the power spectrum analogous to the broad wing of the perturbed gains. For a larger size of perturbation, the effect of the chromatic gain errors become more significant and the chromatic errors along with the strong foreground power observed at the short baselines lead to leakage at very low $k_\perp$ modes. This is consistent with the results of \citet{Orosz2019} in the sense that there is a trend of the foreground leakage at very low and high $\kperp$.

\begin{figure*}[t!]
\centering
\includegraphics[scale=0.5]{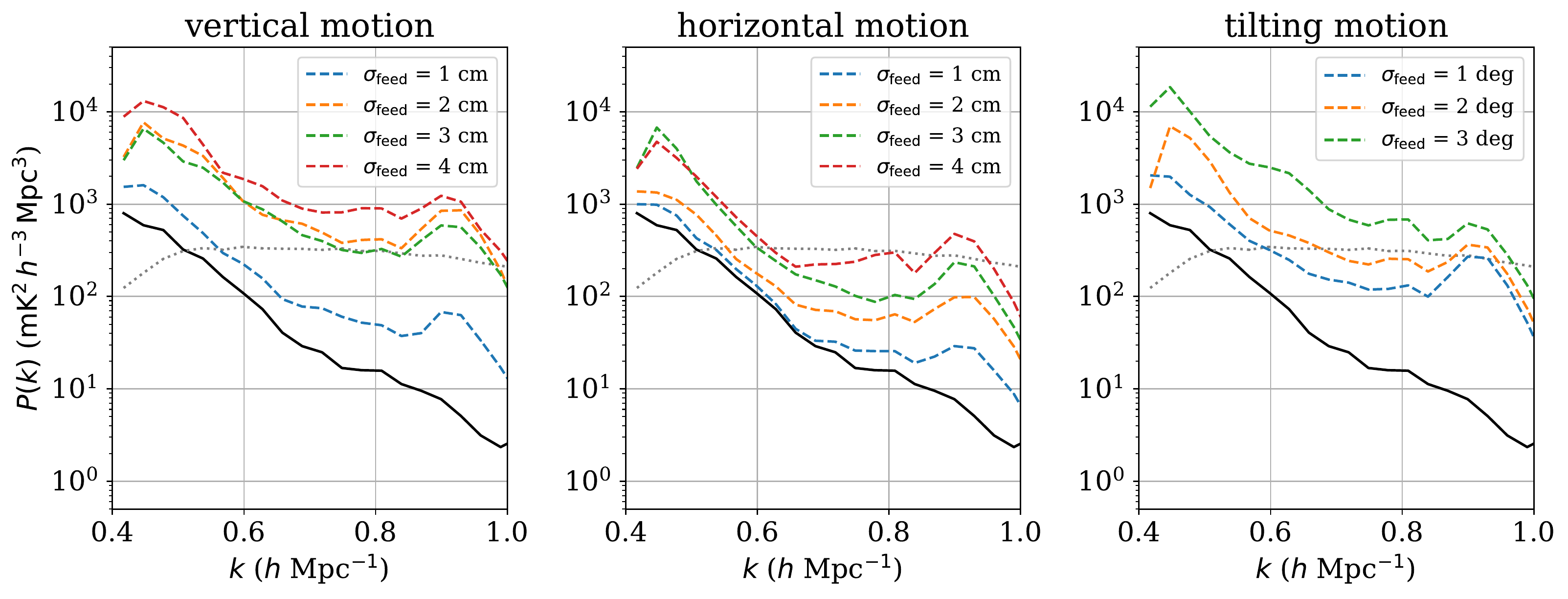}
\caption{Spherically averaged one-dimensional foreground power spectrum estimates for vertical (left), horizontal (middle), and tilting (right) feed motion. The power spectra are constructed from the two-dimensional power spectra (Figure~\ref{fig:pspec_all_no_mitigation}) with $k$ modes satisfying $\kpara \ge k_{\rm \parallel, hor} + 0.4\,\hMpcinv$ to minimize the intrinsic foreground spillovers due to the spectral structure in the primary beam along with the pitchfork effect. The black solid curve is the power spectrum estimate of the fiducial model and colored dashed curves are for the perturbed beams labelled in the legend. The dotted grey curve denotes the EoR power spectrum with the same cut applied.}
\label{fig:1D_pspec}
\end{figure*}

The amount of foreground leakage is different for different type of feed motion. For the vertical and tilting feed motions, the power spectra exhibit stronger evidence of leakage compared to the horizontal motions. We expect this behavior from the large non-redundancy error in visibilities of the vertical and tilting motions as discussed in Section~\ref{sec:mad_vis}. Especially when $\sigma_{\rm feed} = 4$~cm and $\sigma_{\rm feed} = 3^\circ$, we see the low $\kperp$ modes are heavily contaminated by the foregrounds and are not usable for EoR detection. For the horizontal feed motion which shows smaller $\sigma_{\rm red}$, the contours are least pushed away from the fiducial one but noticeable contamination is still allowed at high $\kperp$. The foreground contamination is expected to be reduced by applying the baseline cut-off in calibration and smooth calibration along with the fringe-rate filter, and this will be discussed in the subsequent study.

As shown previously in Figure~\ref{fig:std_cross}, the non-redundancy in visibilities depends not only on the feed motion but also on the sky model. In order to relate the result to the foreground leakage in the power spectrum, we simulated visibilities by perturbing the beam model for one sky model but keeping the fiducial beam model for the other. We then fed the visibility into the calibration pipeline and estimated the power spectrum. The ratio between power spectra, $P_{\rm pGSM}$ (i.e., fiducial GLEAM + perturbed GSM) and $P_{\rm pGLEAM}$ (i.e., perturbed GLEAM + fiducial GSM) is shown in Figure~\ref{fig:pspec_perturb_one_source}.

We consider $\sigma_{\rm feed} = 3$~cm for the vertical (left panel) and horizontal (middle) motions and $\sigma_{\rm feed} = 3^\circ$ for the tilting motion (right). For the vertical motion, the small $P_{\rm pGSM}/P_{\rm pGLEAM}$ is predicted because of predominant $\sigma_{\rm red}$ of the GLEAM over $\sigma_{\rm red}$ of the GSM. This may demonstrate the variation in the width of the main lobe with the vertical feed motion is the key feature inducing the foreground leaking into the EoR window. On the contrary, the larger $P_{\rm pGSM}/P_{\rm pGLEAM}$ for the tilting motion can be explained by the fact that $\sigma_{\rm red}$ of GSM is predominant over $\sigma_{\rm red}$ of GLEAM at short baselines, which may arise from the variation of the side lobes covering the bright galactic plane. The horizontal motion, which has relatively small $\sigma_{\rm red}$ for both GLEAM and GSM compared to other motions, reveals the impact of GSM is greater than that of GLEAM in forming the foreground leakage. This can be understood in the context that the beam-weighted flux density of the GSM can be significantly larger than that of the GLEAM. For example, the flux density of the point sources inside the main lobe (e.g., inside 20$\times$20 degree sq around zenith) is about 1.4 $\times$ 10$^3$ Jy, while that of GSM inside the side lobe (60$^\circ <$ zenith angle $<$ 90$^\circ$) is about 8 $\times$ 10$^5$ Jy. For the feed displacement by 5~cm along the $x$-axis, the change to the power beam area for the main lobe is about 1.4 $\times$ 10$^{-4}$, whereas that for the side lobe is about 2.6 $\times$ 10$^{-5}$. This means the change to the beam-weighted flux density due to the feed motion is $\sim$0.2 Jy for the point sources and is $\sim$20 Jy for the GSM. This implies the contribution of the GSM with the feed motion would be 100 times larger than that of the point source and that's why we see a larger impact of the GSM for the horizontal motion.

A one-dimensional power spectrum is formed based on the two-dimensional power spectrum shown in Figure~\ref{fig:pspec_all_no_mitigation}. We first selected the region least affected by the intrinsic foreground spillovers, even shown in the fiducial model, arising from the chromatic primary beam convolved with the pitchfork effect by setting a constant buffer above the horizon limit, $\kpara \ge k_{\rm \parallel, hor} + k_{\rm \parallel, buffer}$. For a conservative choice to minimize the effect of the spillovers, $k_{\rm \parallel, buffer} = 0.4 \, \hMpcinv$ is taken into account. Though this buffer size is aggressive in terms of removing significant amount of $k$ modes with relatively stronger EoR signals, it enables us to explore the behavior of the foreground leakage due to the feed motions.

We then turned the filtered power spectrum into the spherically averaged one-dimensional power spectrum by averaging cylindrical bins into corresponding spherical bins. Figure~\ref{fig:1D_pspec} shows the results for the vertical (left), horizontal (middle), and tilting (right) feed motions. The fiducial model (black solid line) crosses the EoR power spectrum (grey dotted line) at $k \sim 0.5 \, \hMpcinv$ where $P(k) \sim 300 \, {\rm mK^2} \, h^{-3} \, {\rm Mpc^3}$. Since the EoR power spectrum declines with $k$, we focus on $0.5 < k < 1.0 \,\hMpcinv$ where we can achieve relatively high sensitivity compared to $k > 1.0 \,\hMpcinv$ if thermal noise is included. For the given EoR model, our noiseless simulation shows the foreground power spectrum is smaller than the EoR power spectrum at $k \gtrsim 0.6 \,\hMpcinv$ for the lateral motions except for $\sigma_{\rm feed} =$ 4~cm. For the vertical feed motion, the foreground power spectrum smaller than the EoR power when $\sigma_{\rm feed} = 1$~cm but the EoR power is buried under the foreground power spectrum for other cases. Our choice of tilts also makes the foreground power spectrum similar to or greater than the EoR power spectrum for most cases of the perturbation. We expect the foreground power can be dropped below the EoR level and set a requirement of feed positioning if appropriate mitigation for the chromatic gain errors is applied, which will be explored in Kim et al. (in prep).

\section{Conclusions}
In this study, we have characterized the effects of feed positional perturbations on the electromagnetic properties of the HERA antenna's beam pattern by using the CST's full-wave electromagnetic time-domain solver. Previous studies have so far been limited to studying analytic models of per-antenna beam perturbations for HERA. We significantly extend upon these works by using realistic CST simulations in a full forward-model context, and apply a realistic calibration pipeline to explore its impact on modern data analysis pipelines.

We separate the feed motions into three classes, including vertical, horizontal, and tilting motions. The vertical feed motion was found to mainly drive changes to the width of the main lobe, whereas the horizontal and tilting motions mainly perturb the pointing angle of the main lobe. Relatively larger perturbation in side lobes is observed for the tilting motion than the translation motions.

With the perturbed beams for 320 antennas, we simulated visibility measurements using the GLEAM sky survey and the GSM sky model for point sources and diffuse sky, respectively. Figure~\ref{fig:std_cross} shows different feed motions are responsible for different patterns of non-redundancy in cross-correlation visibilities. The uniform non-redundancy across the antenna separation shown in the vertical feed motion is primarily caused by the response of the point sources to variations in the main lobe, while the high non-redundancy concentrated at short baselines shown in the horizontal and tilting motions arises from the response of the diffuse source to changes of the side lobes.

The introduction of non-redundancies in the visibility due to per-antenna feed motions breaks the assumptions of redundant calibration and thus introduces chromatic errors into the gains. It also imparts gain errors into the absolute calibration step performed after redundant calibration. Fractional gain errors and $\chi^2$ statistics reveal that the chromatic gain errors increase with the size of perturbation in the feed motion. The Fourier transform of the gain solutions shows excess power at delay $\gtrsim 300$~ns compared to the fiducial case, which results in foreground leakage outside the wedge in the power spectrum.

In the two-dimensional power spectrum, the chromatic gain errors cause the foreground to leak from the wedge, which can significantly reduce the accessible size of the EoR window. If the perturbation is small ($\sigma_{\rm feed} \lesssim 1$~cm or $\sigma_{\rm feed} \lesssim 1^\circ$), we found that the foreground power can be suppressed to a level similar to the fiducial case at low $\kperp$ (i.e., short baseline) if we coherently average visibilities over redundant baselines. Since high $\kperp$ modes have less redundant baselines, more foreground leakage is therefore observed relative to low $\kperp$ for all cases of the feed motions. When the feed positions are perturbed more than 1~cm in vertical displacement or 1$^\circ$ in tilt, there are considerable foreground contamination at both low and high $\kperp$ modes due to the chromatic gain errors.

Figure~\ref{fig:pspec_perturb_one_source} demonstrates the foreground leakage is mainly caused by GLEAM point sources in the main lobe of the perturbed beam from the vertical feed motions. We also see that the diffuse GSM component in the side lobes are the main contributors of the leakage in the case of horizontal and tilting feed motions, which is corroborated by Figure~\ref{fig:std_cross}.
 
Based on the spherically averaged one-dimensional power spectrum analysis, $\sigma_{\rm feed} = 1$~cm for the horizontal motion may allow us to retain the EoR window with the least foreground bias. Unlike the lateral feed motion, the vertical and tilting motions introduce more foreground power leakage and the EoR signal is barely above the foreground power spectrum when the feeds are perturbed more than $\sigma_{\rm feed} = 1$~cm or $\sigma_{\rm feed} = 1^\circ$. We expect the stringent feed positioning requirement can be loosened once fine spectral structure in gain solutions can be suppressed using some of the mitigation techniques proposed below.

So far, we have illustrated that antenna feed position offsets can introduce unwanted foreground leakage beyond the horizon limit in the power spectrum. The level of the foreground leakage is expected to be minimized with several different mitigation methods. One major intrinsic foreground spillover, even in the ideal fiducial model, is the pitchfork effect. This pitchfork power leakage level can be reduced by excluding the sky around the horizon, which can be accomplished by filtering out small fringe rates \citep{Parsons2016}. Regarding the leakage associated with chromatic gain errors, including only certain baselines in calibration is a potential solution. For example, \citet{Ewall-wice2017} and \citet{Orosz2019} found down-weighting long baselines in calibration can reduce the chromatic gain errors and the leakage of foreground in the power spectrum. Another approach is to smooth out high-frequency structure in antenna gain solutions \citep[e.g.,][]{Kern2020a}. The effects of these mitigation methods on the power spectrum estimate will be discussed in Kim et al. (in prep).

Although this study focuses on the middle range of the HERA band, it provides a framework to quantify the instrument configuration requirements and systematic errors. A similar approach can be extended for the rest of the HERA band in future work. 

\section*{Acknowledgements}
This material is based upon work supported by the National Science Foundation under Grant Nos. 1636646 and 1836019 and institutional support from the HERA collaboration partners.  This research is funded in part by the Gordon and Betty Moore Foundation through Grant GBMF5212 to the Massachusetts Institute of Technology. The National Radio Astronomy Observatory is a facility of the National Science Foundation operated under cooperative agreement by Associated Universities, Inc. Bang D. Nhan is a Jansky Fellow of the National Radio Astronomy Observatory. Nicholas S. Kern gratefully acknowledges support from the MIT Pappalardo fellowship.

\bibliography{main}{}
\bibliographystyle{aasjournal}

\appendix
\section{Beam interpolation along the feed motion direction}
\label{sec:beam_interpolation}
The primary beams simulated by CST are sampled at regular grid positions, which means the beams interpolated at random feed positions of interest may contain errors arising from the interpolation along the feed motion direction. It is important to evaluate and minimize the potential errors in the interpolation to study the effect of perturbed beams driven by the feed motion.

The interpolation is performed on far-field electric fields expressed in Equation~\eqref{eqn:efield}. Because the far-field electric field consists of two components, $E_\theta$ and $E_\phi$, and each component is complex, the interpolation is carried out 4 times at a given frequency, zenith angle, and azimuthal angle along the feed motion direction using $\texttt{scipy}$ python package \citep{Virtanen2020}.

\begin{figure}[h!]
\centering
\includegraphics[scale=0.4]{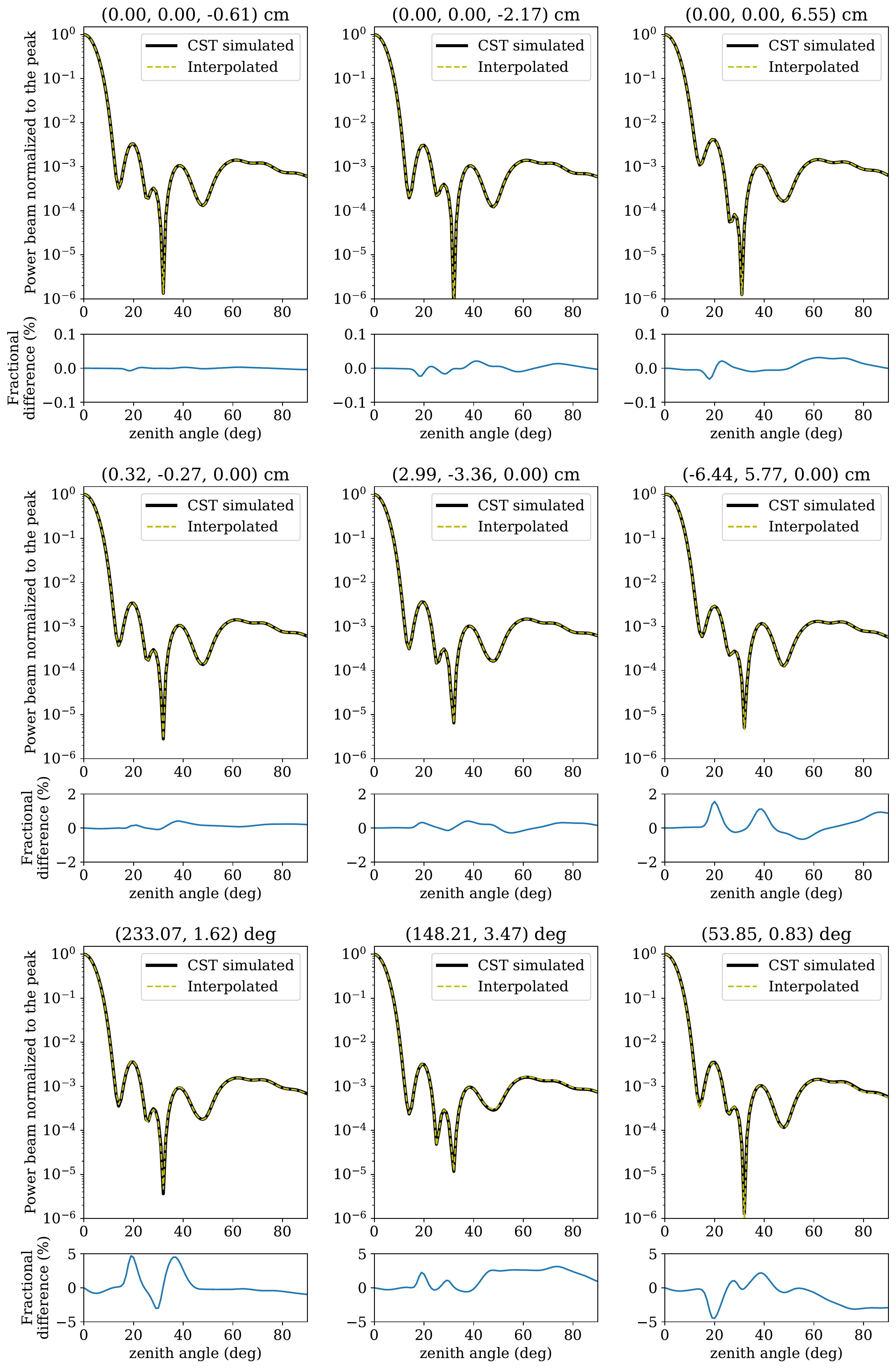}
\caption{Comparison of CST simulated beams and interpolated ones at off-grid feed positions or tilts labeled in the title of each panel at 165 MHz. The title indicates $(x, y, z)$ feed positions for vertical and horizontal feed offsets, and $(\phi, \theta)$ for tilts. Each pair consists of the line profiles of the power beams with zenith angle in the EW direction (top) and their fractional difference (bottom). The top three are for the vertical feed offset, the middle three for the horizontal displacement and the bottom three for the tilting motion.}
\label{fig:interp_beam_all}
\end{figure}

To test the accuracy of the interpolation, we ran CST simulations at randomly chosen feed positions or tilts as noted in the title of each panel of Figure~\ref{fig:interp_beam_all} and compare the CST simulated and interpolated beam. Each pair plot consists of one-dimensional profiles of power beams for the CST simulated and interpolated beams with zenith angle in the EW direction (top) and their fractional difference (bottom). For the pairs in the top row representing the vertical feed offset, the simulated beams and interpolated ones present a good agreement. The fractional difference indicates the interpolation error is less than 0.05\%, which means the error is insignificant compared to the perturbation in the beams due to feed motions. For the horizontal feed motion (pairs in the middle row), the error is about 1\% or less. The errors can be as large as 5\% for the tilts shown in the pair plots in the bottom row but the interpolation error is still less significant than the error of the perturbed beam as shown in Figure~\ref{fig:beam_diff}.

\begin{figure}[h!]
\centering
\includegraphics[scale=0.4]{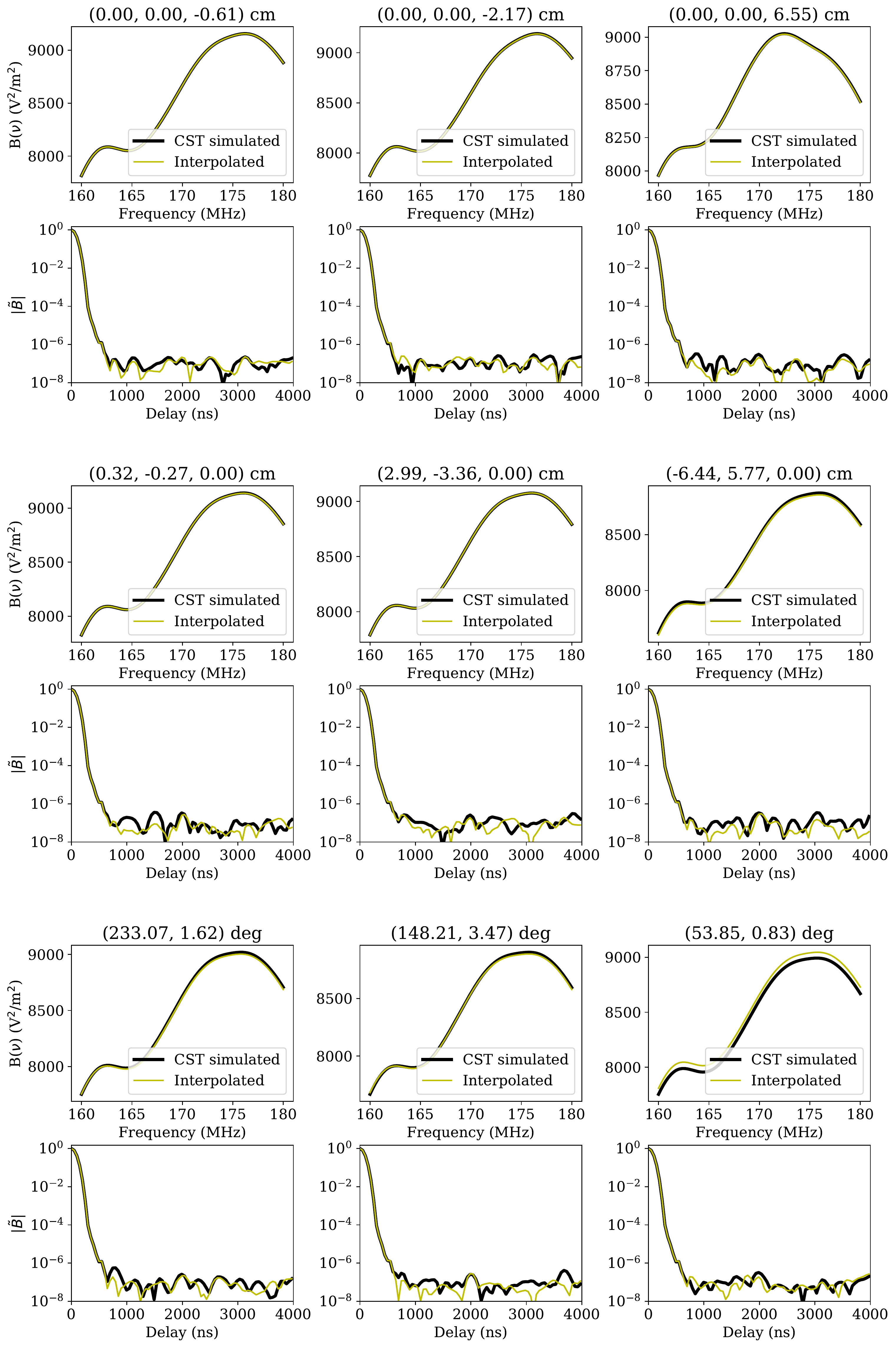}
\caption{Comparison of CST simulated beams and interpolated ones at off-grid feed positions along frequency. The zenith point is chosen for the validation. From top to bottom, each pair plot represents the feed offset in the $z$-axis, in the $xy$ plane, and in tilt. Overall, two power beams in each panel agree better than $\sim$1\%, and the Fourier transform of the interpolated beam does not show extra high-frequency structure compared to that of the CST simulated one.}
\label{fig:interp_beam_all_freqs}
\end{figure}

Figure~\ref{fig:interp_beam_all_freqs} shows comparison between the CST simulated and interpolated beams with frequency at the zenith point. From top to bottom, each pair plot indicates the vertical, horizontal, and tilting feed motions, respectively. Across all panels, the simulated and interpolated beams are well lined up and we found the fractional difference is less than 1\%. In the bottom panel of each pair plot, we show the frequency Fourier transform of the interpolated beam which is consistent with that of the simulated beam, forming the numerical noise floor at around $10^{-7}$. This means there is no additional high-frequency structure introduced by the interpolation. This leaves us with the conclusion that chromatic gain errors and leakage in the power spectrum are mainly caused by the beam error induced by the feed motion rather than the interpolation error.

\bibliographystyle{aasjournal}

\end{document}